\pdfoutput=1
\documentclass[cernpreprint,texlive=2011,txfonts,UKenglish]{atlasdoc}
\usepackage{atlaspackage}
\usepackage{atlasphysics}
\usepackage{atlas-biblatex}

\usepackage{lineno}
\usepackage{subfig}
\usepackage{epsfig}
\usepackage{dcolumn}
\usepackage{graphicx}
\usepackage{xspace}
\usepackage{collref}
\usepackage{url}

\usepackage{hypernat}
\usepackage{heppennames2}
\usepackage{longtable}
\usepackage{multirow}
\usepackage{rotating}

\graphicspath{{logos/}{Figures/}}

\AtlasTitle{Constraints on new phenomena via Higgs boson couplings and invisible decays with the ATLAS detector}
\author{The ATLAS Collaboration}
\AtlasRefCode{HIGG-2015-03}
\PreprintIdNumber{CERN-PH-EP-2015-191}
\AtlasJournal{JHEP}

\AtlasAbstract{
The ATLAS experiment at the LHC has measured the Higgs boson couplings and mass, and searched for invisible Higgs boson decays, using multiple production and decay channels with up to 4.7~fb$^{-1}$ of $pp$ collision data at $\sqrt{s}=7$~TeV and 20.3~fb$^{-1}$ at $\sqrt{s}=8$~TeV.  In the current study, the measured production and decay rates of the observed Higgs boson in the $\gamma\gamma$, $ZZ$, $WW$, $Z\gamma$, $bb$, $\tau\tau$, and $\mu\mu$ decay channels, along with results from the associated production of a Higgs boson with a top-quark pair, are used to probe the scaling of the couplings with mass.  Limits are set on parameters in extensions of the Standard Model including a composite Higgs boson, an additional electroweak singlet, and two-Higgs-doublet models.  Together with the measured mass of the scalar Higgs boson in the $\gamma\gamma$ and $ZZ$ decay modes, a lower limit is set on the pseudoscalar Higgs boson mass of $m_{A}>370$~GeV in the ``hMSSM'' simplified Minimal Supersymmetric Standard Model.  Results from direct searches for heavy Higgs bosons are also interpreted in the hMSSM.  Direct searches for invisible Higgs boson decays in the vector-boson fusion and associated production of a Higgs boson with $W/Z$ ($Z\to \ell\ell$, $W/Z \to jj$) modes are statistically combined to set an upper limit on the Higgs boson invisible branching ratio of 0.25.  The use of the measured visible decay rates in a more general coupling fit improves the upper limit to 0.23, constraining a Higgs portal model of dark matter.
}

\clearpage
\journal{JHEP}
\newcommand{\BRinv}{\ensuremath{\mathrm{BR_{inv}}}}

\newcommand{\ZllhInv}{\ensuremath{Zh \to \ell\ell + E_{{\textrm T}}^{\textrm{miss}}}}
\newcommand{\VBFhInv}{VBF~$\to jj + E_{{\textrm T}}^{\textrm{miss}}$}
\newcommand{\VjjhInv}{\ensuremath{W/Zh \to jj + E_{{\textrm T}}^{\textrm{miss}}}}

\newcommand{\hZZnunununu}{\ensuremath{h \rightarrow ZZ^{*} \rightarrow 4\nu}}
\newcommand{\ETmiss}{\ensuremath{E_{{\textrm T}}^{\textrm{miss}}}}

\newcommand{\scalefac}{s.f.}

\newcommand{\Myggh}{\ensuremath{ggh}}

\DeclareRobustCommand{\Pf}{\HepParticle{f}{}{}\Xspace} % fermion (generic)
\newcommand{\fb}{\ensuremath{\,\mbox{fb}}}

\newcommand{\hgg}{\ensuremath{h{\rightarrow\,}\gamma\gamma}}
\newcommand{\hZg}{\ensuremath{h{\rightarrow\,}Z\gamma}}

\newcommand{\hWWlnln}{\ensuremath{h{\rightarrow\,}WW^{*}{\rightarrow\,}\ell\nu\ell\nu}}

\newcommand{\hZZllll}{\ensuremath{h{\rightarrow\,}ZZ^{*}{\rightarrow\,}4\ell}}

\newcommand{\htt}{\ensuremath{h \rightarrow \tau\tau}}
\newcommand{\hbb}{\ensuremath{h \rightarrow bb}}
\newcommand{\hmm}{\ensuremath{h \rightarrow \mu\mu}}

\def\vec#1{{\mbox{$\boldsymbol{#1}$}}}

\newcommand*{\chisq}{\ensuremath{\raise.4ex\hbox{$\chi$}^{2}}\xspace}
\newcommand{\mT}{\ensuremath{m_{\rm T}}}

\begin{document}

\section{Introduction}{
\label{sec:Intro}

The ATLAS and CMS Collaborations at the Large Hadron Collider (LHC) announced the discovery of a particle consistent with a Higgs boson in 2012~\cite{ATLAS:2012af,Chatrchyan:2012ia}.  Since then, the collaborations have together measured the mass of the particle to be about 125~GeV~\cite{Aad:2014aba, Khachatryan:2014jba, Aad:2015zhl}.  Studies of its spin and parity in bosonic decays have found it to be compatible with a $J^P=0^+$ state~\cite{Aad:2013xqa, Khachatryan:2014kca, Aad:2015mxa}.  Combined coupling fits of the measured Higgs boson production and decay rates within the framework of the Standard Model~(SM) have found no significant deviation from the SM expectations~\cite{Aad:2013wqa, Khachatryan:2014jba, Aad:2015gba}.  These results strongly suggest that the newly discovered particle is indeed a Higgs boson and that a non-zero vacuum expectation value of a Higgs doublet is responsible for electroweak (EW) symmetry breaking~\cite{Englert:1964et,Higgs:1964pj,Guralnik:1964eu}. The observed CP-even Higgs boson is denoted as $h$ throughout this paper.

A crucial question is whether there is only one Higgs doublet, as postulated by the SM, or whether the Higgs sector is more complex, for example with a second doublet leading to more than one Higgs boson of which one has properties similar to those of the SM Higgs boson, as predicted in many theories beyond the Standard Model (BSM).\footnote{The observed CP-even Higgs boson, denoted as $h$ in this paper, is taken to be the lightest Higgs boson, and only heavier additional Higgs bosons are considered.}  The ``hierarchy problem'' regarding the naturalness of the Higgs boson mass, the nature of dark matter, and other open questions that the SM is not able to answer also motivate the search for additional new particles and interactions.  Astrophysical observations provide strong evidence of dark matter that could be explained by the existence of weakly interacting massive particles (see Ref.~\cite{Clowe} and the references therein).  If such decays are kinematically allowed, the observed Higgs boson~\cite{ATLAS:2012af,Chatrchyan:2012ia} might decay to dark matter or other stable or long-lived particles which do not interact significantly with a detector~\cite{Antoniadis:2004se,ArkaniHamed:1998vp,Datta:2004jg,Kanemura:2010sh,Djouadi:2011aa,Djouadi:2012zc}. Such Higgs boson decays are termed ``invisible'' and can be inferred indirectly through final states with large missing transverse momentum. The Higgs boson may also decay to particles that do interact significantly with a detector, such as gluons that produce jets, resulting in final states that cannot be resolved due to the very large backgrounds.  These decays and final states are termed ``undetectable''.

This paper presents searches for deviations from the rates of Higgs boson production and decay predicted by the SM, including both the visible and invisible decay channels, using ATLAS data.  Simultaneous fits of multiple production and decay channels are performed after the removal of overlaps in the event selection of different analyses, and correlations between the systematic uncertainties are accounted for.  The data are interpreted in various benchmark models beyond the SM, providing indirect limits on the BSM parameters.  The limits make different assumptions than those obtained by direct searches for heavy Higgs bosons and invisible Higgs boson decays.

An overview of the experimental inputs is given in Section~\ref{sec:ExpInputs}, and the analysis procedure is described in Section~\ref{sec:Procedure}.  The scaling of the couplings with mass is probed in Section~\ref{sec:MassScaling}.  The measurements of visible Higgs boson decay rates are used to derive limits on model parameters in four representative classes of models:  Minimal Composite Higgs Models (MCHM) in Section~\ref{sec:MCHM}, an additional electroweak singlet in Section~\ref{sec:EWSinglet}, two-Higgs-doublet models (2HDMs) in Section~\ref{sec:2HDM}, and the ``h'' Minimal Supersymmetric Standard Model (hMSSM) in Section~\ref{sec:SUSY}.  The results from direct searches for heavy Higgs bosons are also interpreted in the hMSSM in Section~\ref{sec:SUSY}.  The combination of direct searches for invisible Higgs boson decays is discussed in Section~\ref{sec:HInv}, and the combination of all visible and invisible Higgs boson decay channels is described in Section~\ref{sec:VisibleInvisible}.  This is used together with the visible decays to constrain a Higgs portal model of dark matter in Section~\ref{sec:DarkMatter}.  Finally, Section~\ref{sec:Conclusions} is devoted to the conclusions.

}
\section{Experimental inputs}{
\label{sec:ExpInputs}

For the determination of the couplings in the visible Higgs boson decay channels, the experimental inputs include search results and measurements of Higgs boson decays:  \hgg~\cite{Aad:2014eha}, \hZZllll~\cite{Aad:2014eva}, \hWWlnln~\cite{ATLAS:2014aga, Aad:2015ona}, \hZg~\cite{Aad:2014fia}, \hbb~\cite{Aad:2014xzb}, \htt~\cite{Aad:2015vsa}, and \hmm~\cite{Aad:2014xva} $(\ell=e,\mu$).  Search results from $tth$ associated production with \hgg~\cite{Aad:2014lma}, \hbb~\cite{Aad:2015gra}, and final states with multiple leptons~\cite{Aad:2015iha} are included.  In addition, the constraints on the Higgs boson invisible decay branching ratio use direct searches for Higgs boson decays to invisible particles in events with dileptons or dijets with large missing transverse momentum, \ETmiss.  These inputs include the search for a Higgs boson, produced through vector-boson fusion (VBF) and thus accompanied by dijets, that decays invisibly and results in missing transverse momentum (\VBFhInv)~\cite{Aad:2015txa}; the search for a Higgs boson, which subsequently decays invisibly, produced in association with a $Z$ boson that decays to dileptons (\ZllhInv~\cite{Aad:2014iia}); and the search for a Higgs boson, which afterwards decays invisibly, produced together with a $W$ or $Z$ boson that decays hadronically (\VjjhInv~\cite{Aad:2015uga}).  These searches are based on up to 4.7~fb$^{-1}$ of $pp$ collision data at $\sqrt{s}=7$~TeV and up to 20.3~fb$^{-1}$ at $\sqrt{s}=8$~TeV.

Each measurement or search classifies candidate events into exclusive categories based on the expected kinematic properties of different Higgs boson production processes.  This both improves the sensitivity and enables discrimination between different Higgs boson production modes.  Each search channel is designed to be mostly sensitive to the product of a Higgs boson production cross section and decay branching ratio.  The combination of the visible decay search channels is used~\cite{Aad:2015gba} to determine the couplings of the Higgs boson to other SM particles.  The input analyses, their results, and small changes to them applied for use in this combination are described there.

Direct searches for additional heavy Higgs bosons ($H$, $A$, and $H^{\pm}$) are not used in the fits discussed here, but their results are interpreted in the hMSSM benchmark model for comparison.

}
\section{Analysis procedure}{
\label{sec:Procedure}

In the benchmark models considered, the couplings of the Higgs boson to fermions and vector bosons are modified by functions of the model parameters.  In all cases, it is assumed that the modifications of the couplings do not change the Higgs boson production or decay kinematics significantly.  Thus the expected rate of any given process can be obtained through a simple rescaling of the SM couplings and no acceptance change due to kinematics in each BSM scenario is included.  A simultaneous fit of the measured rates in multiple production and decay modes is used to constrain the BSM model parameters.  The Higgs boson mass was measured by ATLAS to be $m_{h} = 125.36 \pm 0.37$~(stat)~$\pm 0.18$~(syst)~GeV~\cite{Aad:2014aba}.  The best-fit value is used throughout this paper; the uncertainty on the mass is not included.

The statistical treatment of the data is described in Refs.~\cite{paper2012prd,LHC-HCG,Moneta:2010pm,HistFactory,ROOFIT}.  Confidence intervals use the test statistic $t_{\scriptsize \vec{\alpha}} = -2 \ln \Lambda(\vec{\alpha})$, which is based on the profile likelihood ratio~\cite{Cowan:2010st}:
\begin{equation}
  \Lambda(\vec\alpha) = \frac{L\big(\vec\alpha\,,\,\hat{\hat{\vec\theta}}(\vec\alpha)\big)}
{L(\hat{\vec\alpha},\hat{\vec\theta})\label{eq:LH}}\quad .
\end{equation}

The likelihood in Eq.~(\ref{eq:LH}) depends on one or more parameters of interest $\vec\alpha$, such as the Higgs boson production times branching ratio strength $\mu$, the mass $m_h$, and coupling scale factors $\kappa_{i}$.  Systematic uncertainties and their correlations~\cite{paper2012prd} are modelled by introducing nuisance parameters $\vec\theta$ centred at their nominal values.  For the visible decay channels, the treatment of systematic uncertainties is the same as that used in Higgs boson coupling measurements~\cite{Aad:2015gba}.  For the invisible decay channels, the expected event counts for the signals, backgrounds and control regions are taken from Monte Carlo (MC) predictions or data-driven estimations as described in Refs.~\cite{Aad:2015txa,Aad:2014iia,Aad:2015uga}.
The nuisance parameters for each individual source of uncertainty are applied on the relevant expected rates so that the correlated effects of the uncertainties are taken into account.  

The single circumflex in the denominator of Eq.~(\ref{eq:LH}) denotes the unconditional maximum-likelihood estimate of a parameter.  The double circumflex in the numerator denotes the ``profiled'' value, namely the conditional maximum-likelihood estimate for given fixed values of the parameters of interest $\vec\alpha$.

For each production mode $j$ and visible decay channel $k$, $\mu$ is normalised to the SM expectation for that channel so that $\mu=1$ corresponds to the SM Higgs boson hypothesis and $\mu=0$ to the background-only hypothesis:
\begin{equation}
\mu = \frac{\sigma_{j} \times {\rm{BR}}_{k}} {\sigma_{j,{\rm{SM}}} \times {\rm{BR}}_{k,{\rm{SM}}}} \quad ,
\end{equation}
where $\sigma_{j}$ is the production cross section, ${\rm{BR}}_{k}$ is the branching ratio, and the subscript ``SM'' denotes their SM expectations.

For the invisible decay mode, $\mu$ is the production cross section for each production mode $j$ times the invisible decay branching ratio \BRinv, normalised to the total SM rate for the production mode in question:
\begin{equation}
\label{eq:invBR}
\mu = \frac{\sigma_{j}} {\sigma_{j,{\rm{SM}}}} \times \BRinv \quad .
\end{equation}
Thus the SM is recovered at $\mu=0$ when \BRinv$=0$.

Other parameters of interest characterise each particular scenario studied, including the mass scaling parameter $\epsilon$ and the ``vacuum expectation value'' parameter $M$ for the scaling of the couplings with mass (Section~\ref{sec:MassScaling}), compositeness scaling parameter $\xi$ for the Higgs boson compositeness models (Section~\ref{sec:MCHM}), squared coupling ${\kappa^\prime}^2$ of the heavy Higgs boson in the electroweak singlet model (Section~\ref{sec:EWSinglet}), $\cos(\beta-\alpha)$ and $\tan\beta$ for the 2HDM (Section~\ref{sec:2HDM}), pseudoscalar Higgs boson mass $m_{A}$ and $\tan\beta$ for the hMSSM model (Section~\ref{sec:SUSY}), and Higgs boson invisible decay branching ratio \BRinv\ for the studies of Higgs boson invisible decays (Section~\ref{sec:InvHiggsDecays}).

The likelihood function for the Higgs boson coupling measurements is built as a product of the likelihoods of all measured Higgs boson channels, where for each channel the likelihood is built using sums of signal and background probability density functions in the discriminating variables.  These discriminants are chosen to be the $\gamma\gamma$ and $\mu\mu$ mass spectra for \hgg~\cite{Aad:2014eha} and \hmm~\cite{Aad:2014xva} respectively; the transverse mass, $\mT$, distribution\footnote{The transverse mass $\mT$ is defined as:  $\mT = \sqrt{ (E_{\rm T}^{\ell\ell} + p_{\rm T}^{\nu\nu})^2 - | \vec{p}_{\rm T}^{\ell\ell} + \vec{p}_{\rm T}^{\nu\nu}|^2}$, where $E_{\rm T}^{\ell\ell} = \sqrt{(p_{\rm T}^{\ell\ell})^2 + (m_{\ell\ell})^2}$, $\vec{p}_{\rm T}^{\ell\ell}$ ($\vec{p}_{\rm T}^{\nu\nu}$) is the vector sum of the lepton (neutrino) transverse momenta, and $p_{\rm T}^{\ell\ell}$ ($p_{\rm T}^{\nu\nu}$) is its modulus.} for \hWWlnln~\cite{ATLAS:2014aga, Aad:2015ona}; the distribution of a boosted decision tree (BDT) response for \htt~\cite{Aad:2015vsa} and \hbb~\cite{Aad:2014xzb}; the $4\ell$ mass spectrum and a BDT in the \hZZllll\ channel~\cite{Aad:2014eva}; the \ETmiss\ distribution for the \VBFhInv~\cite{Aad:2015txa}, \ZllhInv~\cite{Aad:2014iia}, and \VjjhInv~\cite{Aad:2015uga} channels.  The distributions are derived primarily from MC simulation for the signal, and both the data and simulation contribute to them for the background.

The couplings are parameterised using scale factors denoted $\kappa_{i}$, which are defined as the ratios of the couplings to their corresponding SM values.  The production and decay rates are modified from their SM expectations accordingly, as expected at leading order~\cite{Heinemeyer:2013tqa}.  This procedure is performed for each of the models probed in Sections~\ref{sec:MassScaling}--\ref{sec:InvHiggsDecays}, using the coupling parameterisation given for each model. For example, taking the narrow-width approximation~\cite{Uhlemann:2008pm, Goria:2011wa}, the rate for the process $gg\to h\to ZZ^*\to 4\ell$ relative to the SM prediction can be parameterised~\cite{Heinemeyer:2013tqa} as:
\begin{equation}
\mu = \frac{\sigma\times {\rm BR}}{\sigma_{\rm SM}\times {\rm BR_{\rm SM}}}
    = \frac{\kappa_g^2\cdot\kappa_Z^2}{\kappa_h^2} \quad .
\end{equation}

Here $\kappa_g$ is the scale factor for the loop-induced coupling to the gluon through the top and bottom quarks, where both the top and bottom couplings are scaled by $\kappa_f$, and $\kappa_Z$ is the coupling scale factor for the $Z$ boson.  The scale factor for the total width of the Higgs boson, $\kappa_h^2$, is calculated as a squared effective coupling scale factor.  It is defined as the sum of squared coupling scale factors for all decay modes, $\kappa_j^2$, each weighted by the corresponding SM partial decay width $\Gamma_{jj}^\text{SM}$~\cite{Heinemeyer:2013tqa}:
\begin{equation}
\kappa_{h}^2 = \sum\limits_{jj} \frac{\kappa_j^2 \Gamma_{jj}^\text{SM}}
{\Gamma_{h}^\text{SM}} \quad ,
\label{eq:CH2_def}
\end{equation}
where $\Gamma_{h}^\text{SM}$ is the SM total width and the summation runs over $WW$, $ZZ$, $\gamma\gamma$, $Z\gamma$, $gg$, $tt$, $bb$, $cc$, $ss$, $\tau\tau$, and $\mu\mu$.  The present experimental sensitivity to Higgs boson decays to charm and strange quarks with the current data is very low.  Therefore the scale factors of the corresponding couplings are taken to be equal to those of the top and bottom quarks, respectively, which have the same quantum numbers.  The couplings to the first-generation quarks (up and down) and the electron are negligible.

In most of the models considered (Sections~\ref{sec:MassScaling}--\ref{sec:SUSY}), it is assumed that no new production or decay modes beyond those in the SM are kinematically open.  In addition, the production or decays through loops are resolved in terms of the contributing particles in the loops, taking non-negligible contributions only from SM particles.  For example, the $W$ boson provides the dominant contribution to the $h\to\gamma\gamma$ decay (followed by the top quark), such that the effective coupling scale factor $\kappa_\gamma$ is given by:
\begin{eqnarray}
\kappa_{\gamma}^2(\kappa_{b}, \kappa_{t}, \kappa_{\tau}, \kappa_{W}) &=& \frac{\sum\limits_{i,j (i \ge j)}\kappa_i \kappa_j\cdot\Gamma_{\gamma\gamma}^{i j}}{\sum\limits_{i,j (i \ge j)}\Gamma_{\gamma\gamma}^{ij}} \quad ,
\label{eq:CgammaNLOQCD}
\end{eqnarray}
where $\Gamma_{\gamma\gamma}^{i j}$ is the contribution to the diphoton decay width due to a particle loop ($i=j$) or due to the interference between two particles ($i \ne j$), and where the summations run over the $W$ boson, top and bottom quarks, and tau lepton.  Contributions from other charged particles in the SM are negligible.  The destructive interference between the $W$ and top loops, as well as the contributions from other charged particles in the loops, are thus accounted for. Similarly, for the loop-induced $h \to Z\gamma$ and $gg \to h$ processes the effective coupling scale factors are given by:
\begin{eqnarray}
\kappa_{Z\gamma}^2(\kappa_{b}, \kappa_{t}, \kappa_{\tau}, \kappa_{W}) &=& \frac{\sum\limits_{i,j (i \ge j)}\kappa_i \kappa_j\cdot\Gamma_{Z\gamma}^{i j}}{\sum\limits_{i,j (i \ge j)}\Gamma_{Z\gamma}^{ij}}
\end{eqnarray}
\begin{eqnarray}
 \kappa_{g}^2(\kappa_{b}, \kappa_{t}) &=& \frac{\kappa_{t}^2\cdot\sigma_{\Myggh}^{tt} +\kappa_{b}^2\cdot\sigma_{\Myggh}^{bb} +\kappa_{t}\kappa_{b}\cdot\sigma_{\Myggh}^{tb}}{\sigma_{\Myggh}^{tt}+\sigma_{\Myggh}^{bb}+\sigma_{\Myggh}^{tb}} \label{eq:CgNLOQCD} \quad ,
\end{eqnarray}
where $\sigma_{\Myggh}^{tt}$, $\sigma_{\Myggh}^{bb}$, $\sigma_{\Myggh}^{tb}$ are the respective contributions to the gluon fusion cross section from a top loop, bottom loop, and the interference of the top and bottom loops.

In the searches for  Higgs boson decays to invisible particles discussed in Section~\ref{sec:InvHiggsDecays}, it is assumed that there are no new production modes beyond the SM ones; however, the possibility of new decay modes is left open.  The couplings associated with Higgs boson production and decays through loops are not resolved, but rather left as effective couplings.

Confidence intervals are extracted by taking $t_{\scriptsize \vec{\alpha}}$ to follow an asymptotic \chisq distribution with the corresponding number of degrees of freedom~\cite{Cowan:2010st}.  For the composite Higgs boson (see Section~\ref{sec:MCHM}), EW singlet (Section~\ref{sec:EWSinglet}), and invisible Higgs boson decays (Section~\ref{sec:InvHiggsDecays}), a physical boundary imposes a lower bound on the model parameter under study.  The confidence intervals reported are based on the profile likelihood ratio where parameters are restricted to the allowed region of parameter space, as in the case of the $\tilde{t}_{\mu}$ test statistic described in Ref.~\cite{Cowan:2010st}. This restriction of the likelihood ratio to the allowed region of parameter space is similar to the Feldman--Cousins technique~\cite{PhysRevD.57.3873} and provides protection against artificial exclusions due to fluctuations into the unphysical regime.  However, the confidence interval is defined by the standard \chisq cutoff, leading to overcoverage near the physical boundaries as demonstrated by toy examples.  The Higgs boson couplings also have physical boundaries in the two-dimensional parameter space of the 2HDM (see Section~\ref{sec:2HDM}) and hMSSM (Section~\ref{sec:SUSY}) models, which are treated in a similar fashion.

For the combination of the direct searches for invisible Higgs boson decays, confidence intervals in \BRinv\ are defined using the CL$_{S}$ procedure~\cite{CLs_2002} in order to be consistent with the convention used in the individual searches.  For the constraints on \BRinv\ from the rate measurements in visible Higgs boson decay channels, and from the overall combination of visible and invisible decay channels, the log-likelihood ratio is used in order to be consistent with the convention used in deriving the Higgs boson couplings via the combination of visible decay channels.

Table~\ref{tab:Couplings} summarises the relevant best-fit value, interval at the 68\% confidence level (CL), and/or upper limit at the 95\% CL for physical quantities of interest.  These include the overall signal strength, the scale factors for the Higgs boson couplings and total width, and the Higgs boson invisible decay branching ratio in various parameterisations.  The BSM models probed with these parameters are also indicated.  The overall signal strength measured is above 1.  The extracted coupling scale factors can be similar to or less than 1 because the measured rate for \hbb, which has a branching ratio of 57\% in the SM for $m_{h}=125.36$~GeV, is lower than (although still compatible with) the expected rate.

\setlength{\tabcolsep}{12pt}
\begin{table*}[!htbp]
\small
\begin{tabular}{lp{0.15\linewidth}|c|c|c}
\hline\hline
\multicolumn{2}{c|}{Model} & \multirow{2}{*}{\begin{minipage}{1.5cm}Coupling\\Parameter\end{minipage}} & Description & Measurement \\
& & & & \\
\hline\hline
\multirow{5}{*}{1} & \multirow{5}{*}{\begin{minipage}{2.9cm}Mass scaling\\ parameterisation\end{minipage}} & $\kappa_Z$ & \begin{minipage}{3.6cm}$Z$ boson coupling~\scalefac\end{minipage} & $[-1.06, -0.82] \cup [0.84, 1.12]$ \\
  \cline{3-5}
  & & $\kappa_W$ & \begin{minipage}{3.6cm}$W$ boson coupling~\scalefac\end{minipage} & $0.91 \pm 0.14$ \\
  \cline{3-5}
  & & $\kappa_t$ & \begin{minipage}{3.6cm}$t$-quark coupling~\scalefac\end{minipage} & $0.94 \pm 0.21$ \\
  \cline{3-5}
  & & $\kappa_b$ & \begin{minipage}{3.6cm}$b$-quark coupling~\scalefac\end{minipage} & $[-0.90, -0.33] \cup [0.28, 0.96]$ \\
  \cline{3-5}
  & & $\kappa_{\tau}$ & \begin{minipage}{3.6cm}Tau lepton coupling~\scalefac\end{minipage} & $[-1.22, -0.80] \cup [0.80, 1.22]$ \\
  \cline{3-5}
  & & $\kappa_{\mu}$ & \begin{minipage}{3.6cm}Muon coupling~\scalefac\end{minipage} & $<2.28$ at 95\% CL \\
\hline
\multirow{2}{*}{2} & \multirow{2}{*}{\begin{minipage}{2.9cm}MCHM4,\\ EW singlet\end{minipage}} & \multirow{2}{*}{$\mu_{h}$} & \multirow{2}{*}{\begin{minipage}{3.6cm}Overall signal strength\end{minipage}} & \multirow{2}{*}{$1.18^{+0.15}_{-0.14}$} \\
& & & & \\
\hline
\multirow{4}{*}{3} & \multirow{4}{*}{\begin{minipage}{2.9cm}MCHM5,\\ 2HDM Type I\end{minipage}} & \multirow{2}{*}{$\kappa_V$} & \multirow{2}{*}{\begin{minipage}{3.6cm}Vector boson ($W$, $Z$)\\ coupling~\scalefac\end{minipage}} & \multirow{2}{*}{$1.09 \pm 0.07$} \\
& & & & \\
\cline{3-5}
 & & \multirow{2}{*}{$\kappa_F$} & \multirow{2}{*}{\begin{minipage}{3.6cm}Fermion ($t$, $b$, $\tau$, \ldots) \\coupling~\scalefac\end{minipage}} & \multirow{2}{*}{$1.11 \pm 0.16$} \\
& & & & \\
\hline
\multirow{10}{*}{4} & \multirow{10}{*}{\begin{minipage}{2.9cm}2HDM Type II,\\hMSSM\end{minipage}} & \multirow{3}{*}{$\lambda_{Vu} = \kappa_{V}/\kappa_{u}$} & \multirow{3}{*}{\begin{minipage}{3.6cm}Ratio of vector boson to \\up-type fermion ($t$, $c$, \ldots) \\coupling~\scalefac\end{minipage}} & \multirow{3}{*}{$0.92^{+0.18}_{-0.16}$} \\
& & & & \\
& & & & \\
\cline{3-5}
  & & \multirow{3}{*}{$\kappa_{uu} = \kappa_{u}^{2}/\kappa_{h}$} & \multirow{3}{*}{\begin{minipage}{3.6cm}Ratio of squared up-type\\ fermion coupling~\scalefac\ to \\total width~\scalefac\end{minipage}} & \multirow{3}{*}{$1.25 \pm 0.33$} \\
& & & & \\
& & & & \\
\cline{3-5}
  & & \multirow{4}{*}{$\lambda_{du} = \kappa_{d} / \kappa_{u}$} & \multirow{4}{*}{\begin{minipage}{3.6cm}Ratio of down-type \\fermion ($b$, $\tau$, \ldots) to \\up-type fermion \\coupling~\scalefac\end{minipage}} & \multirow{4}{*}{$[-1.08, -0.81] \cup [0.75, 1.04]$} \\
& & & & \\
& & & & \\
& & & & \\
\hline
\multirow{9}{*}{5} & \multirow{9}{*}{\begin{minipage}{2.9cm}2HDM\\ Lepton-specific\end{minipage}} & \multirow{3}{*}{$\lambda_{Vq} = \kappa_{V}/\kappa_{q}$} & \multirow{3}{*}{\begin{minipage}{3.6cm}Ratio of vector boson to \\quark ($t$, $b$, \ldots) \\coupling~\scalefac\end{minipage}} & \multirow{3}{*}{$1.03^{+0.18}_{-0.15}$} \\
& & & & \\
& & & & \\
\cline{3-5}
  & & \multirow{3}{*}{$\kappa_{qq} = \kappa_{q}^{2}/\kappa_{h}$} & \multirow{3}{*}{\begin{minipage}{3.6cm}Ratio of squared quark \\ coupling~\scalefac\ to total width~\scalefac\end{minipage}} & \multirow{3}{*}{$1.03^{+0.24}_{-0.20}$} \\
& & & & \\
& & & & \\
\cline{3-5}
  & & \multirow{2}{*}{$\lambda_{\ell q} = \kappa_{\ell} / \kappa_{q}$} & \multirow{2}{*}{\begin{minipage}{3.6cm}Ratio of lepton ($\tau$, $\mu$, $e$) \\ to quark coupling~\scalefac\end{minipage}} & \multirow{2}{*}{$[-1.34, -0.94] \cup [0.94, 1.34]$} \\
& & & & \\
\hline
\multirow{15}{*}{6} & \multirow{11}{*}{\begin{minipage}{2.9cm}Higgs portal\\ (Baseline config.\\ of vis. \& inv.\\ Higgs boson\\ decay channels:\\ general coupling \\ param., no \\ assumption about \\ $\kappa_{W,Z}$)\end{minipage}} & $\kappa_{Z}$ & \begin{minipage}{3.6cm}$Z$ boson coupling~\scalefac\end{minipage} & $0.99 \pm 0.15$ \\
\cline{3-5}
  & & $\kappa_{W}$ & \begin{minipage}{3.6cm}$W$ boson coupling~\scalefac\end{minipage} & $0.92 \pm 0.14$ \\
\cline{3-5}
  & & \multirow{2}{*}{$\kappa_t$} & \multirow{2}{*}{\begin{minipage}{3.6cm}$t$-quark coupling~\scalefac\end{minipage}} & \multirow{2}{*}{$1.26^{+0.32}_{-0.34}$} \\
& & & & \\
  \cline{3-5}
  & & $\kappa_b$ & \begin{minipage}{3.6cm}$b$-quark coupling~\scalefac\end{minipage} & $0.61 \pm 0.28$  \\
  \cline{3-5}
  & & \multirow{2}{*}{$\kappa_{\tau}$} & \multirow{2}{*}{\begin{minipage}{3.6cm}Tau lepton coupling~\scalefac\end{minipage}} & \multirow{2}{*}{$0.98^{+0.20}_{-0.18}$} \\
& & & & \\
  \cline{3-5}
  & & $\kappa_{\mu}$ & \begin{minipage}{3.6cm}Muon coupling~\scalefac\end{minipage} & $<2.25$ at 95\% CL \\
  \cline{3-5}
  & & \multirow{2}{*}{$\kappa_{g}$} & \multirow{2}{*}{\begin{minipage}{3.6cm}Gluon coupling~\scalefac\end{minipage}} & \multirow{2}{*}{$0.92^{+0.18}_{-0.15}$} \\
& & & & \\
  \cline{3-5}
  & & \multirow{2}{*}{$\kappa_{\gamma}$} & \multirow{2}{*}{\begin{minipage}{3.6cm}Photon coupling~\scalefac\end{minipage}} & \multirow{2}{*}{$0.90^{+0.16}_{-0.14}$} \\
& & & & \\
  \cline{3-5}
  & & $\kappa_{Z\gamma}$ & \begin{minipage}{3.6cm}$Z\gamma$ coupling~\scalefac\end{minipage} & $<3.15$ at 95\% CL \\
  \cline{3-5}
  & & \BRinv & \begin{minipage}{3.6cm}
Invisible branching ratio\end{minipage} & $<0.23$ at 95\% CL \\
\hline\hline
\end{tabular}
\caption{Measurements of the overall signal strength, scale factors (\scalefac) for the Higgs boson couplings and total width, and the Higgs boson invisible decay branching ratio, in different coupling parameterisations, along with the BSM models or parameterisations they are used to probe.  The measurements quoted for Models 1--5 were derived in Ref.~\cite{Aad:2015gba}, while those for Model~6 are derived in this paper.  The production modes are taken to be the same as those in the SM in all cases.  In Models~1--3, decay modes identical to those in the SM are taken. For Models~4--5, the coupling parameterisations and measurements listed do not require such an assumption, which is however made when deriving limits on the underlying parameters of these BSM models.  No assumption about the total width is made for Model~6.}
\label{tab:Couplings}
\end{table*}

}

\section{Mass scaling of couplings}{
\label{sec:MassScaling}

The observed rates in different channels are used to determine how the Higgs boson couplings to other particles scale with the masses of those particles.  The measurements~\cite{Aad:2015gba} of the scale factors for the couplings of the Higgs boson to the $Z$ boson, $W$ boson, top quark, bottom quark, $\tau$ lepton, and muon -- namely [$\kappa_{Z}$, $\kappa_{W}$, $\kappa_{t}$, $\kappa_{b}$, $\kappa_{\tau}$, $\kappa_{\mu}$] -- are given in Model~1 of Table~\ref{tab:Couplings}.  The coupling scale factors to different species of fermions and vector bosons, respectively, are expressed in terms of the parameters [$\epsilon$, $M$]~\cite{Ellis:2013lra}, where $\epsilon$ is a mass scaling parameter and $M$ is a ``vacuum expectation value'' parameter whose SM value is $v \approx 246$~GeV:
\begin{equation}
\begin{array}{lcl}
\vspace{0.2cm}
\kappa_{F,i} = & v\frac{m_{F,i}^\epsilon}{M^{1+\epsilon}} \\
\kappa_{V,j} = & v\frac{m_{V,j}^{2\epsilon}}{M^{1+2\epsilon}} \quad ,
\end{array}
\end{equation}
where $m_{F,i}$ denotes the mass of each fermion species (indexed $i$) and $m_{V,j}$ denotes each vector-boson mass (indexed $j$). The mass scaling of the couplings, as well as the vacuum expectation value, of the SM are recovered with parameter values $\epsilon=0$ and $M=v$, which produce $\kappa_{F,i}=\kappa_{V,j}=1$.  The value $\epsilon=-1$ would correspond to light Higgs boson couplings that are independent of the particle mass.

Combined fits to the measured rates are performed with the mass scaling factor $\epsilon$ and the vacuum expectation value parameter $M$ as the two parameters of interest.  Figure~\ref{fig:Epsilon_M} shows contours of the two-dimensional likelihood as a function of $\epsilon$ and $M$.  The measured and expected values from one-dimensional likelihood scans are given in Table~\ref{tab:MassScaling}.  The mass scaling of the couplings in the SM ($\epsilon=0$) is compatible with the data within one~std.~dev.  The extracted value of $\epsilon$ is close to 0, indicating that the measured couplings to fermions and vector bosons are consistent with the linear and quadratic mass dependence, respectively, predicted in the SM.  The best-fit value for $M$ is less than $v \approx 246$~GeV because the measured overall signal strength $\mu_{h}$ is greater than 1, with the data being compatible with the SM within about 1.5~std.~dev.

\begin{figure}[tbp!]
\begin{center}
\includegraphics[width=0.6\textwidth]{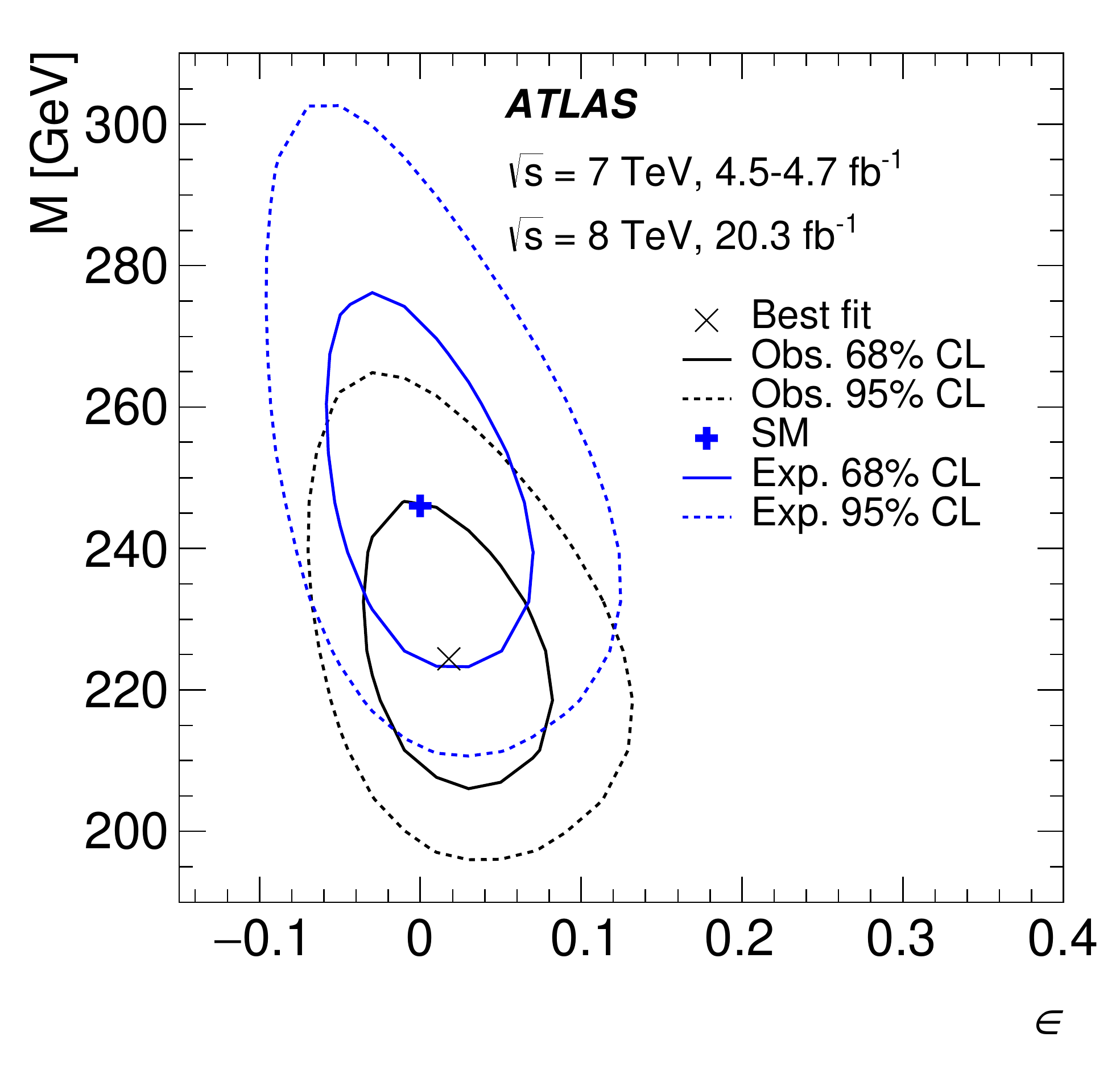}
\caption{Two-dimensional confidence regions as a function of the mass scaling factor $\epsilon$ and the vacuum expectation value parameter $M$.  The likelihood contours where $-2\ln\Lambda=2.3$ and $-2\ln\Lambda=6.0$, corresponding approximately to the 68\% CL (1~std.~dev.) and the 95\% CL (2~std.~dev.) respectively, are shown for both the data and the prediction for a SM Higgs boson.  The best fit to the data and the SM expectation are indicated as $\times$ and $+$ respectively.
}
\label{fig:Epsilon_M}
\end{center}
\end{figure}

\begin{table}[tbp]
\begin{center}
\begin{tabular}{c|cccc}
\hline
\hline
Parameter & \multicolumn{2}{c}{Obs.} & \multicolumn{2}{c}{Exp.} \\
\hline
$\epsilon$ & \multicolumn{2}{c}{$0.018 \pm 0.039$} & \multicolumn{2}{c}{$0.000 \pm 0.042$} \\
$M$ & \multicolumn{2}{c}{$224^{+14}_{-12}$~GeV} & \multicolumn{2}{c}{$246^{+19}_{-16}$~GeV} \\
\hline
\hline
\end{tabular}
\caption{Observed and expected measurements of the mass scaling parameter $\epsilon$ and the ``vacuum expectation value'' parameter $M$.}
\label{tab:MassScaling}
\end{center}
\end{table}

}
\section{Minimal composite Higgs model}{
\label{sec:MCHM}

Minimal Composite Higgs Models (MCHM)~\cite{Agashe:2004rs, PhysRevD.75.055014,Carena:2007ua,DeCurtis:2011yx,Marzocca:2012zn,Carmi:2012in, Panico:2015jxa} represent a possible explanation for the scalar naturalness problem, wherein the Higgs boson is a composite, pseudo-Nambu--Goldstone boson rather than an elementary particle.  In such cases, the Higgs boson couplings to vector bosons and fermions are modified with respect to their SM expectations as a function of the Higgs boson compositeness scale, $f$.  Corrections due to new heavy resonances such as vector-like quarks~\cite{Aguilar-Saavedra:2013qpa} are taken to be sub-dominant.  Production or decays through loops are resolved in terms of the contributing particles in the loops, assuming only contributions from SM particles.  It is assumed that there are no new production or decay modes besides those in the SM.

The MCHM4 model~\cite{Agashe:2004rs} is a minimal SO(5)/SO(4) model where the SM fermions are embedded in spinorial representations of SO(5).  Here the ratio of the predicted coupling scale factors to their SM expectations, $\kappa$, can be written in the particularly simple form:
\begin{equation}
\begin{array}{lcl}
\vspace{0.2cm}
\kappa = \kappa_V = \kappa_F = \sqrt{1-\xi} \quad , \\
\end{array}
\label{eqn:MCHM_Type4}
\end{equation}
where $\xi = v^2 / f^2$ is a scaling parameter (with $v$ being the SM vacuum expectation value) such that the SM is recovered in the limit $\xi \to 0$, namely $f \to \infty$.  The combined signal strength, $\mu_{h}$, which is equivalent to the coupling scale factor, $\kappa = \sqrt{\mu_{h}}$, was measured using the combination of the visible decay channels~\cite{Aad:2015gba} and is listed in Model~2 of Table~\ref{tab:Couplings}.  The experimental measurements are interpreted in the MCHM4 scenario by rescaling the rates in different production and decay modes as functions of the coupling scale factors $\kappa = \kappa_V = \kappa_F$, taking the same production and decay modes as in the SM.  This is done in the same way as described in Section~\ref{sec:Procedure}.  The coupling scale factors are in turn expressed as functions of $\xi$ using Eq.~(\ref{eqn:MCHM_Type4}).

Figure~\ref{fig:MCHM_1D}(a) shows the observed and expected likelihood scans of the Higgs compositeness scaling parameter, $\xi$, in the MCHM4 model.  This model contains a physical boundary restricting to $\xi \ge 0$, with the SM Higgs boson corresponding to $\xi=0$.  Ignoring this boundary, the scaling parameter is measured to be $\xi = 1 - \mu_{h} = -0.18 \pm 0.14$, while the expectation for the SM Higgs boson is $0 \pm 0.14$.  The best-fit value observed for $\xi$ is negative because $\mu_{h}>$1 is measured.  The statistical and systematic uncertainties are of similar size.  Accounting for the lower boundary produces an observed (expected) upper limit at the 95\% CL of $\xi <$ 0.12 (0.23), corresponding to a Higgs boson compositeness scale of $f>710$~GeV (510~GeV).  The observed limit is stronger than expected because $\mu_{h} >$1 was measured~\cite{Aad:2015gba}.

\begin{figure}[tbp]
\centering
\begin{center}
\vspace{-1cm}
\subfloat[MCHM4]{\includegraphics[width=0.45\textwidth]{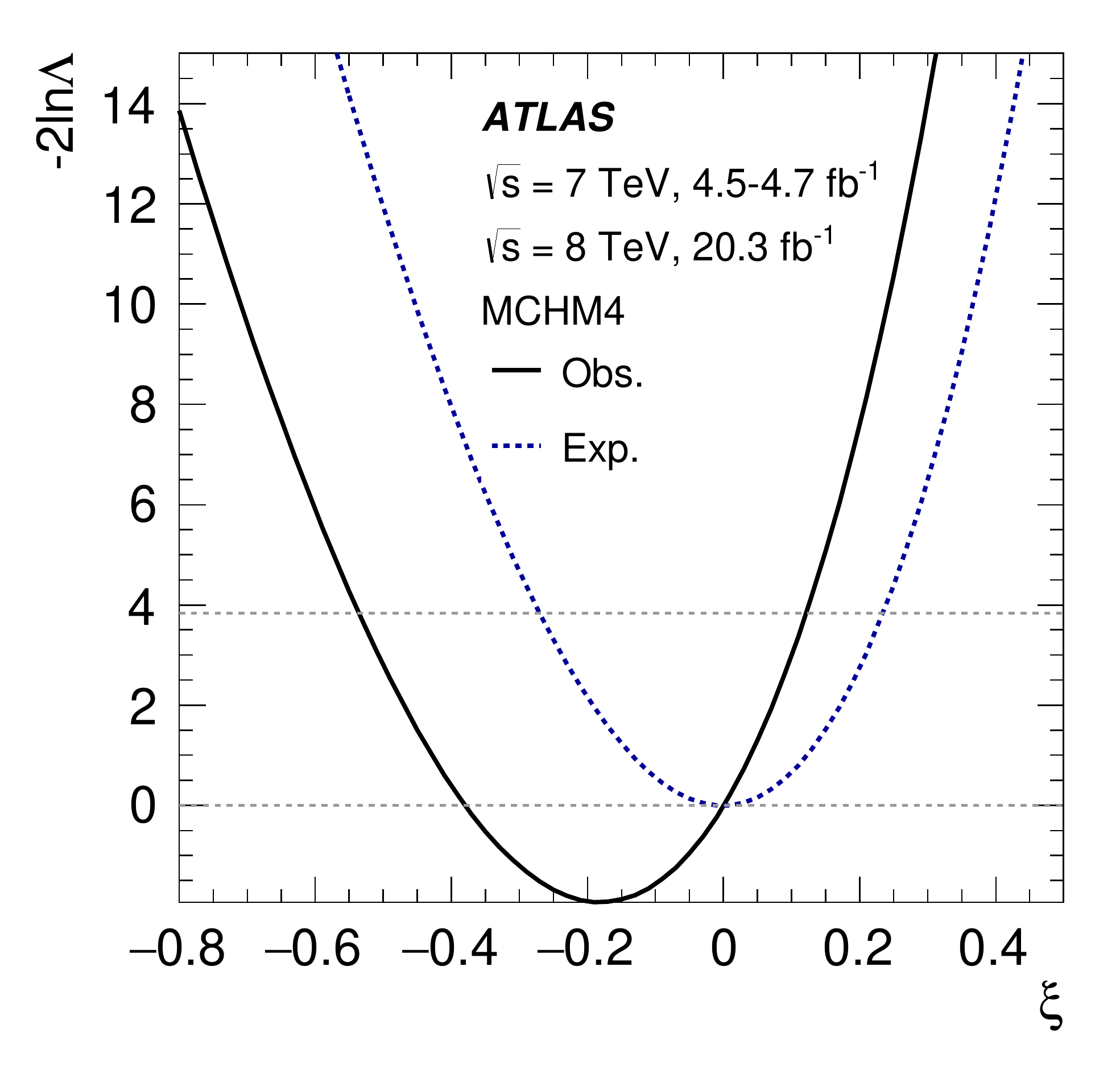}}
\subfloat[MCHM5]{\includegraphics[width=0.45\textwidth]{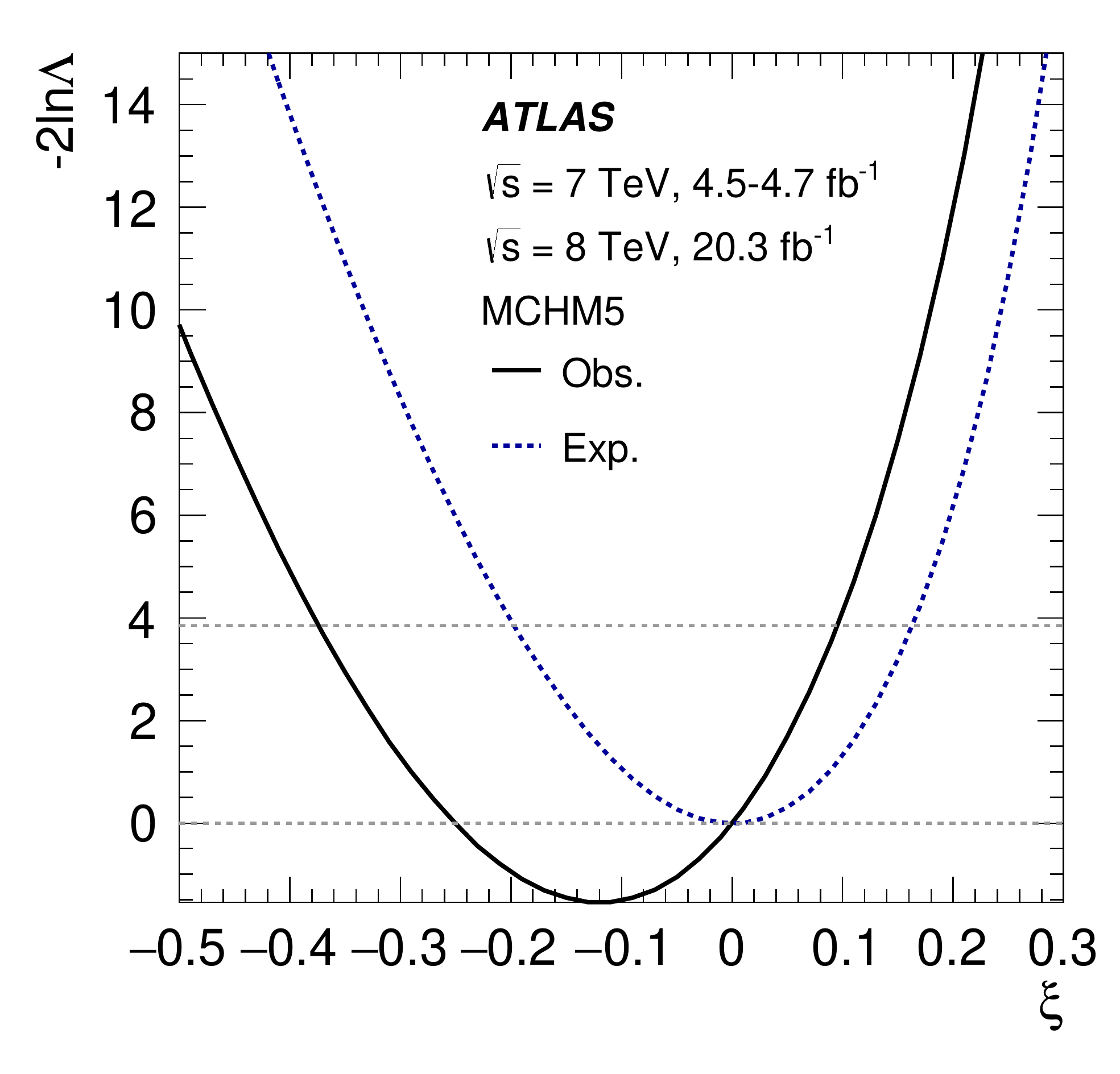}} \\
\caption{Observed (solid) and expected (dashed) likelihood scans of the Higgs compositeness scaling parameter, $\xi$, in the MCHM4 and MCHM5 models.  The expected curves correspond to the SM Higgs boson.  The line at $-2\ln\Lambda=0$ corresponds to the most likely value of $\xi$ within the physical region $\xi \ge 0$.  The line at $-2\ln\Lambda=3.84$ corresponds to the one-sided upper limit at approximately the 95\% CL (2~std.~dev.), given $\xi \ge 0$.}
\label{fig:MCHM_1D}
\end{center}
\end{figure}

Similarly, the MCHM5 model~\cite{PhysRevD.75.055014, Carena:2007ua} is an SO(5)/SO(4) model where the SM fermions are embedded in fundamental representations of SO(5).  Here the measured rates are expressed in terms of $\xi$ by rewriting the coupling scale factors [$\kappa_{V}$, $\kappa_{F}$] as:
\begin{equation}
\begin{array}{lcl}
\vspace{0.2cm}
\kappa_V = \sqrt{1-\xi} \\
\kappa_F = \frac{1-2\xi}{\sqrt{1-\xi}} \quad , \\
\end{array}
\label{eqn:MCHM_Type5}
\end{equation}
where $\xi = v^2 / f^2$.  The measurements of $\kappa_{V}$ and $\kappa_{F}$~\cite{Aad:2015gba} are given in Model~3 of Table~\ref{tab:Couplings}.  The likelihood scans of $\xi$ in MCHM5 are shown in Figure~\ref{fig:MCHM_1D}(b).  As with the MCHM4 model, the MCHM5 model contains a physical boundary restricting to $\xi \ge 0$, with the SM Higgs boson corresponding to $\xi=0$.  Ignoring this boundary, the composite Higgs boson scaling parameter is determined to be $\xi = -0.12 \pm 0.10$, while $0.00 \pm 0.10$ is expected for the SM Higgs boson.  As above, the best-fit value for $\xi$ is negative because $\mu_{h} >$1 is measured.  Accounting for the boundary produces an observed (expected) upper limit at the 95\% CL of $\xi <$ 0.10 (0.17), corresponding to a Higgs boson compositeness scale of $f>780$~GeV (600~GeV).

Figure~\ref{fig:MCHM} shows the two-dimensional likelihood for a measurement of the vector boson ($\kappa_V$) and fermion ($\kappa_F$) coupling scale factors in the $[\kappa_V,\kappa_F]$ plane, overlaid with predictions as parametric functions of $\xi$ for the MCHM4 and MCHM5 models~\cite{Azatov:2012bz,Espinosa:2012ir,Carmi:2012yp}.  Table~\ref{tab:MCHM} summarises the lower limits at the 95\% CL on the Higgs boson compositeness scale in these models.

\begin{figure}[tbp!]
\begin{center}
\includegraphics[width=0.7\textwidth]{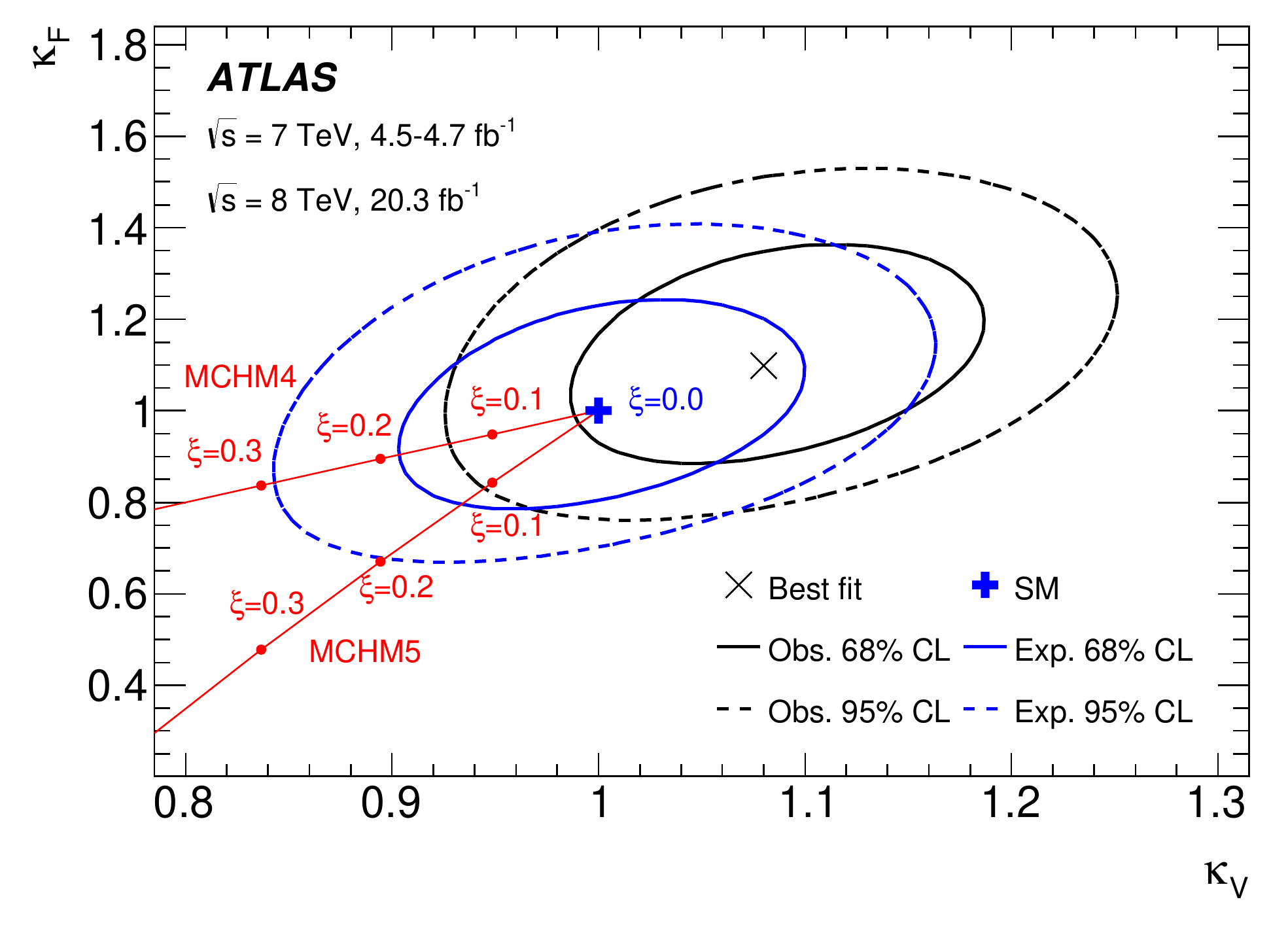}
\caption{Two-dimensional likelihood contours in the $[\kappa_{V},\kappa_{F}]$ coupling scale factor plane, where $-2\ln\Lambda=2.3$ and $-2\ln\Lambda=6.0$ correspond approximately to the 68\% CL (1~std.~dev.) and the 95\% CL (2~std.~dev.), respectively.  The coupling scale factors predicted in the MCHM4 and MCHM5 models are shown as parametric functions of the Higgs boson compositeness parameter $\xi = v^2 / f^2$.  The two-dimensional likelihood contours are shown for reference and should not be used to estimate the exclusion for the single parameter $\xi$.}
\label{fig:MCHM}
\end{center}
\end{figure}

\begin{table}[tbp!]
\begin{center}
\begin{tabular}{c|cc}
\hline
\hline
Model & \multicolumn{2}{c}{Lower limit on $f$} \\
 & Obs. & Exp. \\
\hline
MCHM4 & 710~GeV & 510~GeV \\
MCHM5 & 780~GeV & 600~GeV \\
\hline
\hline
\end{tabular}
\caption{Observed and expected lower limits at the 95\% CL on the Higgs boson compositeness scale $f$ in the MCHM4 and MCHM5 models.}
\label{tab:MCHM}
\end{center}
\end{table}

}

\section{Additional electroweak singlet}{
\label{sec:EWSinglet}

A simple extension to the SM Higgs sector involves the addition of one scalar EW singlet field~\cite{Hill:1987ea,Veltman:1989vw,Binoth:1996au,Schabinger:2005ei,Patt:2006fw,Robens:2015gla,Heinemeyer:2013tqa} to the doublet Higgs field of the SM, with the doublet acquiring a non-zero vacuum expectation value.  This spontaneous symmetry breaking leads to mixing between the singlet state and the surviving state of the doublet field, resulting in two CP-even Higgs bosons, where $h$ ($H$) denotes the lighter (heavier) of the pair.  The two Higgs bosons, $h$ and $H$, are taken to be non-degenerate in mass.  Their couplings to fermions and vector bosons are similar to those of the SM Higgs boson, but each with a strength reduced by a common scale factor, denoted by $\kappa$ for $h$ and $\kappa^\prime$ for $H$.  The coupling scale factor $\kappa$ ($\kappa^\prime$) is the sine (cosine) of the $h$--$H$ mixing angle, so:
\begin{equation}
\kappa^2 +{\kappa^\prime}^2=1 \quad .
\label{eqn:EWSinglet_Uni}
\end{equation}

The lighter Higgs boson $h$ is taken to be the observed Higgs boson.  It is assumed to have the same production and decay modes as the SM Higgs boson does,\footnote{The decays of the heavy Higgs bosons to the light Higgs boson, for example $H \to hh$, are assumed to contribute negligibly to the light Higgs boson production rate.  The contamination from heavy Higgs boson decays (such as $H \to WW$) in light Higgs boson signal regions ($h \to WW$) is also taken to be negligible.} with only SM particles contributing to loop-induced production or decay modes.  In this model, its production and decay rates are modified according to:
\begin{equation}
\begin{array}{lcl}
\vspace{0.2cm}
\sigma_{h}    & = & \kappa^2\times \sigma_{h, \rm SM} \\
\vspace{0.2cm}
\Gamma_{h}     & = & \kappa^2\times \Gamma_{h, \rm SM} \\
{\rm BR}_{h,i}  & = & {\rm BR}_{h, i, {\rm SM}} \quad , \\
\end{array}
\label{eqn:EWSinglet_h}
\end{equation}
where $\sigma_{h}$ denotes the production cross section, $\Gamma_{h}$ denotes the total decay width, BR$_{h,i}$ denotes the branching ratio to the different decay modes $i$, and SM denotes their respective values in the Standard Model.

For the heavier Higgs boson $H$, new decay modes such as $H\to hh$ are possible if they are kinematically allowed.  In this case, the production and decay rates of the $H$ boson are modified with respect to those of a SM Higgs boson with equal mass by the branching ratio of all new decay modes, ${\rm BR}_{H, \rm new}$, as:
\begin{equation}
\begin{array}{lcl}
\vspace{0.2cm}
\sigma_{H}  & = & {\kappa^\prime}^2\times \sigma_{H, \rm SM} \\
\vspace{0.2cm}
\Gamma_{H}  & = & \dfrac{{\kappa^\prime}^2}{1-{\rm BR}_{H, \rm new}}\times \Gamma_{H, \rm SM} \\
{\rm BR}_{H,i} & = & (1-{\rm BR}_{H, \rm new})\times {\rm BR}_{H, {\rm SM}, i} \quad .
\end{array}
\label{eqn:EWSinglet_H}
\end{equation}
Here $\sigma_{H, \rm SM}$, $\Gamma_{H, \rm SM}$, and ${\rm BR}_{H, {\rm SM}, i}$ denote the cross section, total width, and branching ratio for a given decay mode (indexed $i$) predicted for a SM Higgs boson with mass $m_{H}$.

Consequently the overall signal strengths, namely the ratio of production and decay rates in the measured channels relative to the expectations for a SM Higgs boson with corresponding mass, are given by:
\begin{equation}
\begin{array}{lcccl}
\mu_{h}  & = & \cfrac{\sigma_{h}\times {\rm BR}_{h}}{\left(\sigma_{h}\times {\rm BR}_{h}\right)_{\rm SM}} & = & \kappa^2 \\ \\
\mu_{H} & = & \cfrac{\sigma_{H}\times {\rm BR}_{H}}{\left(\sigma_{H}\times {\rm BR}_{H}\right)_{\rm SM}} & = & \kappa'^2\left(1-{\rm BR}_{H, \rm new}\right)  \quad ,
\end{array}
\label{eqn:EWSinglet_mu}
\end{equation}
for $h$ and $H$ respectively, assuming the narrow-width approximation such that interference effects are negligible.

Combining Eqs.~(\ref{eqn:EWSinglet_Uni}) and~(\ref{eqn:EWSinglet_mu}), the squared coupling scale factor of the heavy Higgs boson can be expressed in terms of the signal strength of the light Higgs boson as:
\begin{equation}
{\kappa^\prime}^2 = 1 - \mu_{h} \quad .
\label{eqn:EWSinglet_kPrime}
\end{equation}
This equation for the squared coupling scale factor takes the same form as Eq.~(\ref{eqn:MCHM_Type4}), so the same parameter constraints are expected.

In particular, accounting for the lower boundary yields an observed (expected) upper limit at the 95\% CL of $\kappa'^{2} <$ 0.12 (0.23), which is indicated in Table~\ref{tab:EWSinglet}.
From Eq.~(\ref{eqn:EWSinglet_mu}), this corresponds to the maximum signal strength for contamination by heavy Higgs boson decays in the light Higgs boson signal.  Figure~\ref{fig:ews} shows the limits in the $[\mu_{H}, {\rm BR}_{H, \rm new}]$ plane of the heavy Higgs boson.  Contours of the scale factor for the total width, $\Gamma_{H}/\Gamma_{H, \rm SM}$, based on Eqs.~(\ref{eqn:EWSinglet_H}) and~(\ref{eqn:EWSinglet_mu}), are also illustrated.  These parameters are interesting as potential experimental observables in direct searches for heavy Higgs bosons.  These results are independent of the mass and ${\rm BR}_{H, \rm new}$ of the heavy Higgs boson.

\begin{figure}[tbp!]
\begin{center}
\includegraphics[width=0.7\textwidth]{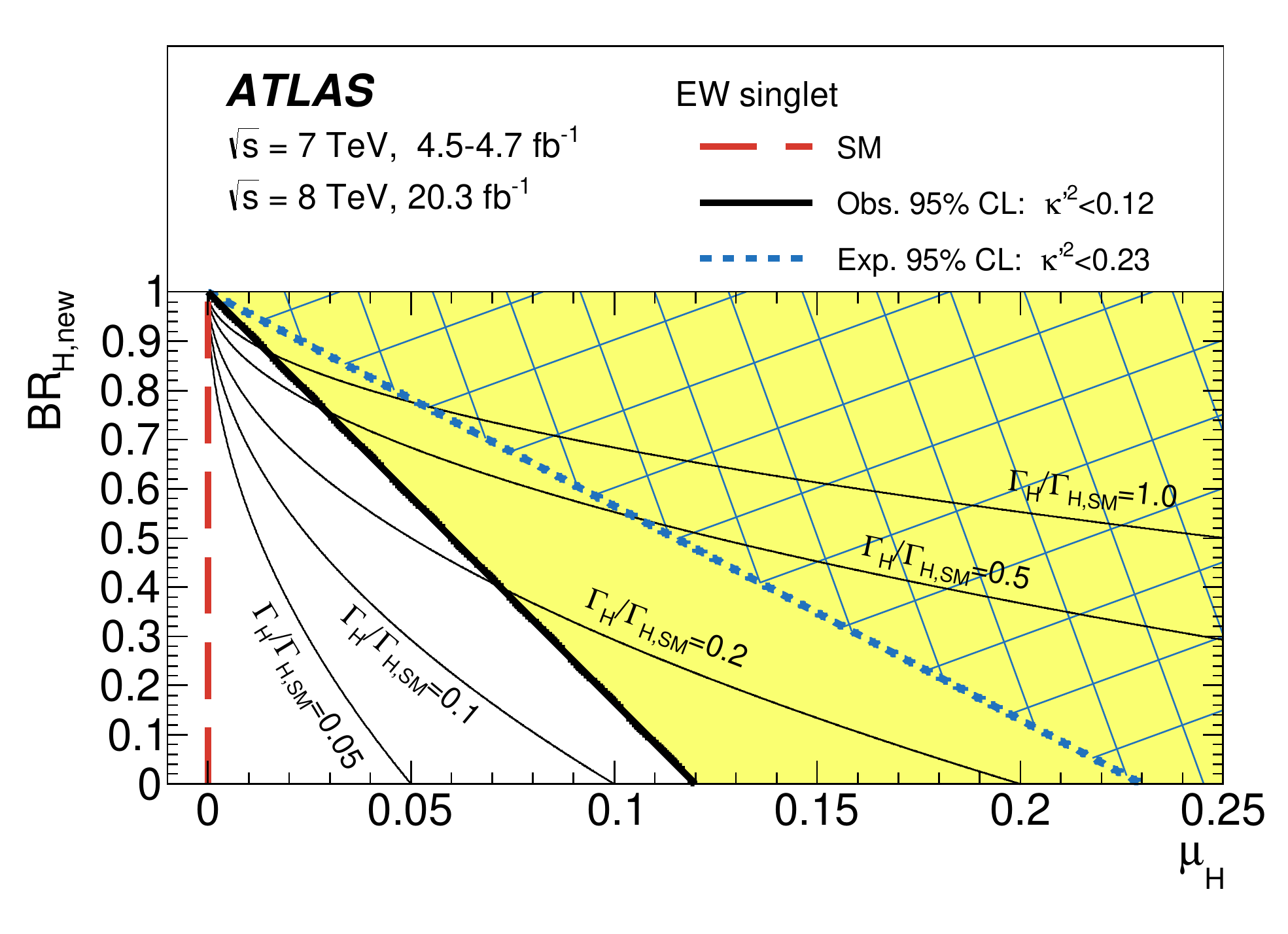}
\caption{Observed and expected upper limits at the 95\% CL on the squared coupling scale factor, $\kappa'^{2}$, of a heavy Higgs boson arising through an additional EW singlet, shown in the $[\mu_{H}, {\rm BR}_{H, \rm new}]$ plane.  The light shaded and hashed regions indicate the observed and expected exclusions, respectively.  Contours of the scale factor for the total width, $\Gamma_{H}/\Gamma_{H, \rm SM}$, of the heavy Higgs boson are also illustrated based on Eqs.~(\ref{eqn:EWSinglet_H}) and~(\ref{eqn:EWSinglet_mu}).}
\label{fig:ews}
\end{center}
\end{figure}

\begin{table}[tbp]
\begin{center}
\begin{tabular}{cc}
\hline
\hline
\multicolumn{2}{c}{Upper limit on ${\kappa^\prime}^2$} \\
Obs. & Exp. \\
\hline
0.12 & 0.23 \\
\hline
\hline
\end{tabular}
\caption{Observed and expected upper limits at the 95\% CL on the squared coupling scale factor of the heavy Higgs boson, ${\kappa^\prime}^2$, in a model with an additional electroweak singlet.}
\label{tab:EWSinglet}
\end{center}
\end{table}}

\section{Two Higgs doublet model} {
\label{sec:2HDM}

Another simple extension to the SM Higgs sector is the 2HDM~\cite{Lee:1973iz,Gunion:2002zf,Branco:2011iw,Heinemeyer:2013tqa}, in which the SM Higgs sector is extended by an additional doublet of the complex field.  Five Higgs bosons are predicted in the 2HDM:  two neutral CP-even bosons $h$ and $H$, one neutral CP-odd boson $A$, and two charged bosons $H^\pm$.  The most general 2HDMs predict CP-violating Higgs boson couplings as well as tree-level flavour-changing neutral currents.  Because the latter are strongly constrained by existing data, the models considered have additional requirements imposed, such as the Glashow--Weinberg condition~\cite{PhysRevD.15.1958, PhysRevD.15.1966}, in order to evade existing experimental bounds.

Both Higgs doublets acquire vacuum expectation values, $v_{1}$ and $v_{2}$ respectively.  Their ratio is denoted by $\tan\beta\equiv v_2/v_1$, and they satisfy $v_{1}^{2}+v_{2}^{2} = v^{2} \approx (246\ {\rm GeV})^2$.  The Higgs sector of the 2HDM can be described by six parameters:  four Higgs boson masses ($m_h$, $m_H$, $m_A$, and $m_{H^\pm}$), $\tan\beta$, and the mixing angle $\alpha$ of the two neutral, CP-even Higgs states.  Gauge invariance fixes the couplings of the two neutral, CP-even Higgs bosons to vector bosons relative to their SM values to be:
\begin{equation}
\begin{array}{lcl}
\vspace{0.2cm}
  g^{\rm 2HDM}_{h VV} / g_{h VV}^{\rm SM} & = & \sin(\beta-\alpha) \\
  g^{\rm 2HDM}_{H VV} / g_{H VV}^{\rm SM} & = & \cos(\beta-\alpha) \quad .\\
\end{array}
\end{equation}
Here $V=W, Z$ and $g_{hVV,HVV}^{\rm SM}$ denote the SM Higgs boson couplings to vector bosons.
\vspace{0.2cm}

The Glashow--Weinberg condition is satisfied by four types of 2HDMs~\cite{Branco:2011iw}:
\begin{itemize}
  \item Type I:  One Higgs doublet couples to vector bosons, while the other couples to fermions.  The first doublet is ``fermiophobic'' in the limit that the two Higgs doublets do not mix.
  \item Type II:  This is an ``MSSM-like'' model, in which one Higgs doublet couples to up-type quarks and the other to down-type quarks and charged leptons.  This model is realised in the Minimal Supersymmetric Standard Model (MSSM) (see Section~\ref{sec:SUSY}).
  \item Lepton-specific:  The Higgs bosons have the same couplings to quarks as in the Type I model and to charged leptons as in Type II.
  \item Flipped:  The Higgs bosons have the same couplings to quarks as in the Type II model and to charged leptons as in Type I.
\end{itemize}
Table~\ref{tab:2HDM} expresses the scale factors for the light Higgs boson couplings, [$\kappa_V$, $\kappa_u$, $\kappa_d$, $\kappa_{\ell}$], in terms of $\alpha$ and $\tan\beta$ for each of the four types of 2HDMs~\cite{Craig:2012pu}.  The coupling scale factors are denoted $\kappa_V$ for the $W$ and $Z$ bosons, $\kappa_u$ for up-type quarks, $\kappa_d$ for down-type quarks, and $\kappa_{\ell}$ for charged leptons.

\setlength{\tabcolsep}{6pt}
\begin{table*}[tb]
\begin{tabular}{c|cccc}
\hline\hline
Coupling scale factor & Type I & Type II & Lepton-specific & Flipped \\
\hline
\hline
$\kappa_V$ & \multicolumn{4}{c}{$\sin(\beta - \alpha)$} \\
\hline
$\kappa_u$ & \multicolumn{4}{c}{$\cos(\alpha) / \sin(\beta)$} \\
\hline
$\kappa_d$ & $\cos(\alpha) / \sin(\beta)$ & $- \sin(\alpha) / \cos(\beta)$ & $\cos(\alpha) / \sin(\beta)$   & $- \sin(\alpha) / \cos(\beta)$ \\
\hline
$\kappa_{\ell}$ & $\cos(\alpha) / \sin(\beta)$ & $- \sin(\alpha) / \cos(\beta)$ & $- \sin(\alpha) / \cos(\beta)$ & $\cos(\alpha) / \sin(\beta)$ \\
\hline\hline
\end{tabular}
\caption{Couplings of the light Higgs boson $h$ to weak vector bosons ($\kappa_{V}$), up-type quarks ($\kappa_{u}$), down-type quarks ($\kappa_{d}$), and charged leptons ($\kappa_{\ell}$), expressed as ratios to the corresponding SM predictions in 2HDMs of various types.}
\label{tab:2HDM}
\end{table*}

The Higgs boson rate measurements in different production and decay modes are interpreted in each of these four types of 2HDMs, taking the observed Higgs boson to be the light CP-even neutral Higgs boson $h$. This is done by rescaling the production and decay rates as functions of the coupling scale factors [$\kappa_V$, $\kappa_u$, $\kappa_d$, $\kappa_{\ell}$].  The measurements of these coupling scale factors or ratios of them~\cite{Aad:2015gba}, taking the same production and decay modes as in the SM, are given in Models~3--5 of Table~\ref{tab:Couplings}.  These coupling scale factors are in turn expressed as a function of the underlying parameters, the two angles $\beta$ and $\alpha$, using the relations shown in Table~\ref{tab:2HDM}.  Here the decay modes are taken to be the same as those of the SM Higgs boson.

After rescaling by the couplings, the predictions agree with those obtained using the \texttt{SUSHI~1.1.1}~\cite{Harlander:2012pb} and \texttt{2HDMC~1.5.1}~\cite{Eriksson:2009ws} programs, which calculate Higgs boson production and decay rates respectively in two-Higgs-doublet models.  The rescaled gluon fusion (ggF) rate agrees with the \texttt{SUSHI} prediction to better than a percent, and the rescaled decay rates show a similar level of agreement.  The cross section for $bbh$ associated production is calculated using \texttt{SUSHI} and included as a correction that scales with the square of the Yukawa coupling to the $b$-quark, assuming that it produces differential distributions that are the same as those in ggF.  The correction is a small fraction of the total production rate for the regions of parameter space where the data would be compatible with the SM at the 95\% CL.

The two parameters of interest correspond to the quantities $\cos(\beta-\alpha)$ and $\tan\beta$.  The 2HDM possesses an ``alignment limit'' at $\cos(\beta-\alpha)=0$~\cite{Branco:2011iw} in which all the Higgs boson couplings approach their respective SM values.  The 2HDM also allows for limits on the magnitudes of the various couplings that are similar to the SM values, but with a negative relative sign of the couplings to particular types of fermions.  These limits appear in the regions where $\cos(\beta+\alpha)=0$, as shown in Table~\ref{tab:2HDM}.  For example, in the Type II model the region where $\cos(\beta+\alpha)=0$, corresponding to the sign change $\alpha \to -\alpha$, has a ``wrong-sign Yukawa limit''~\cite{Ferreira:2014naa,Ferreira:2014dya} with couplings similar to the SM values except for a negative coupling to down-type quarks.  The case for the Flipped model is similar, but with a negative coupling to both the leptons and down-type quarks.  An analogous ``symmetric limit''~\cite{Ferreira:2014dya} appears in the Lepton-specific model.

Figure~\ref{fig:2HDM_Overlay} shows the regions of the $[\cos(\beta-\alpha), \tan\beta]$ plane that are excluded at a CL of at least 95\% for each of the four types of 2HDMs, overlaid with the exclusion limits expected for the SM Higgs sector.  The $\alpha$ and $\beta$ parameters are taken to satisfy $0 \le \beta \le \pi/2$ and $0 \le \beta-\alpha \le \pi$ without loss of generality.  The observed and expected exclusion regions in $\cos(\beta-\alpha)$ depend on the particular functional dependence of the couplings on $\beta$ and $\alpha$, which are different for the down-type quarks and leptons in each of the four types of 2HDMs, as shown in Table~\ref{tab:2HDM}.  There is a physical boundary $\kappa_{V} \le 1$ in all four 2HDM types, to which the profile likelihood ratio is restricted.  The data are consistent with the alignment limit at $\cos(\beta-\alpha)=0$, where the light Higgs boson couplings approach the SM values, within approximately one~std.~dev. or better in each of the models.

In each of the Type~II, Lepton-specific, and Flipped models, at the upper right of the $[\cos(\beta-\alpha), \tan\beta]$ plane where $\tan\beta$ is moderate, there is a narrow, curved region or ``petal'' of allowed parameter space with the surrounding region being excluded.  These three allowed upper petals correspond respectively to an inverted sign of the coupling to down-type fermions (tau lepton and bottom quark), leptons ($\tau$ and $\mu$), or the bottom quark.  These couplings are measured with insufficient precision to be excluded.  There is no upper petal at high $\tan\beta$ in Type~I as all the Yukawa couplings are identical.

In each of the four 2HDM types a similar petal is possible at the lower right of the $[\cos(\beta-\alpha), \tan\beta]$ plane.  For the Type~I, Type~II, Lepton-specific, and Flipped models, this lower petal corresponds respectively to an inverted coupling to fermions, up-type quarks, all quarks, and lastly the up-type quarks and leptons.  In all four cases, the lower petal is rejected since an inverted top quark coupling sign is disfavoured.  The top quark coupling is extracted primarily through its dominant effect in ggF Higgs production, as well as by resolving the Higgs boson decays to diphotons, with one contribution being from the top quark.

\begin{figure*}[tbp]
\centering
\begin{center}
\vspace{-1cm}
\subfloat[Type I]{\includegraphics[width=0.45\textwidth]{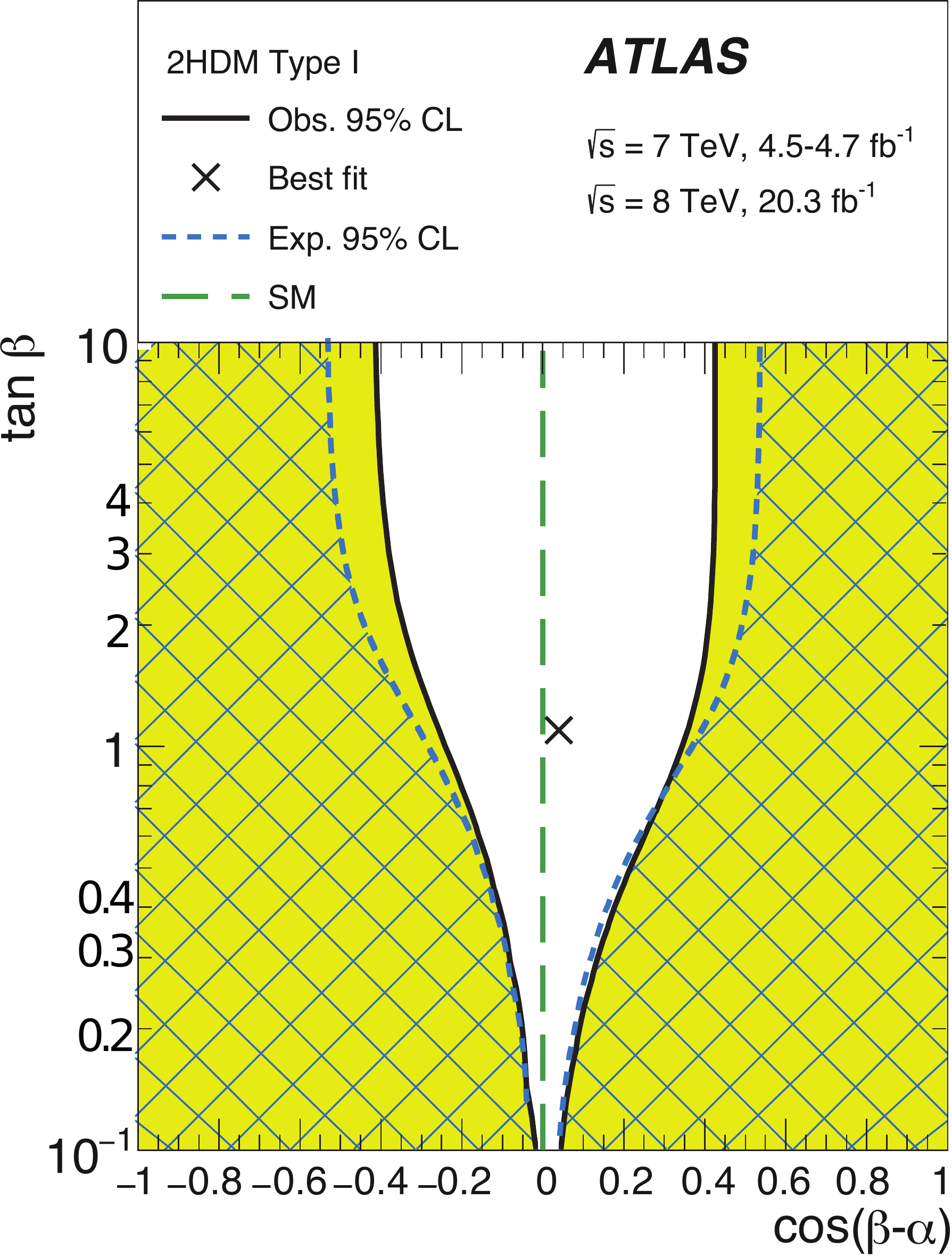}} \hspace{0.75cm}
\subfloat[Type II]{\includegraphics[width=0.45\textwidth]{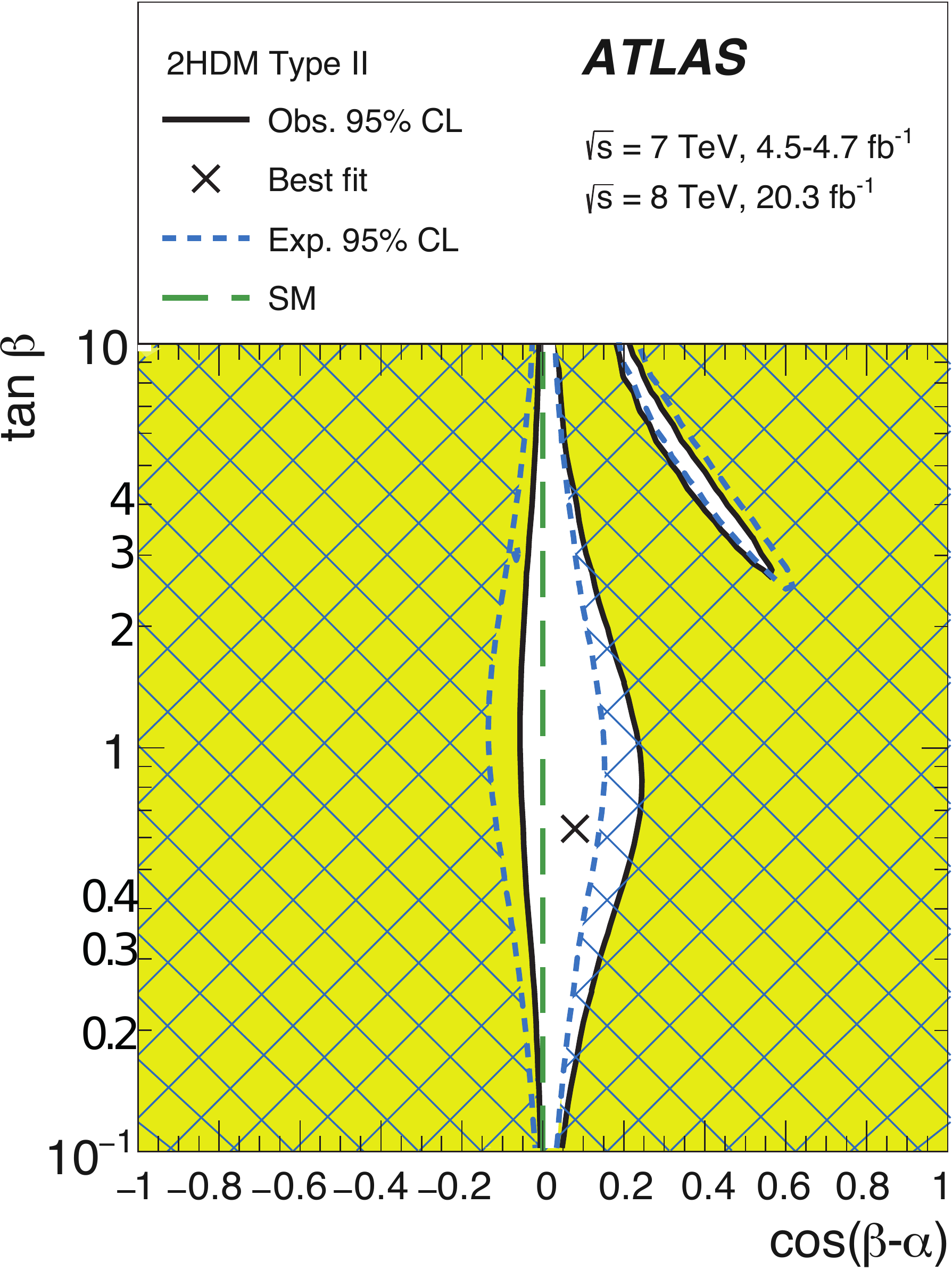}} \\
\subfloat[Lepton-specific]{\includegraphics[width=0.45\textwidth]{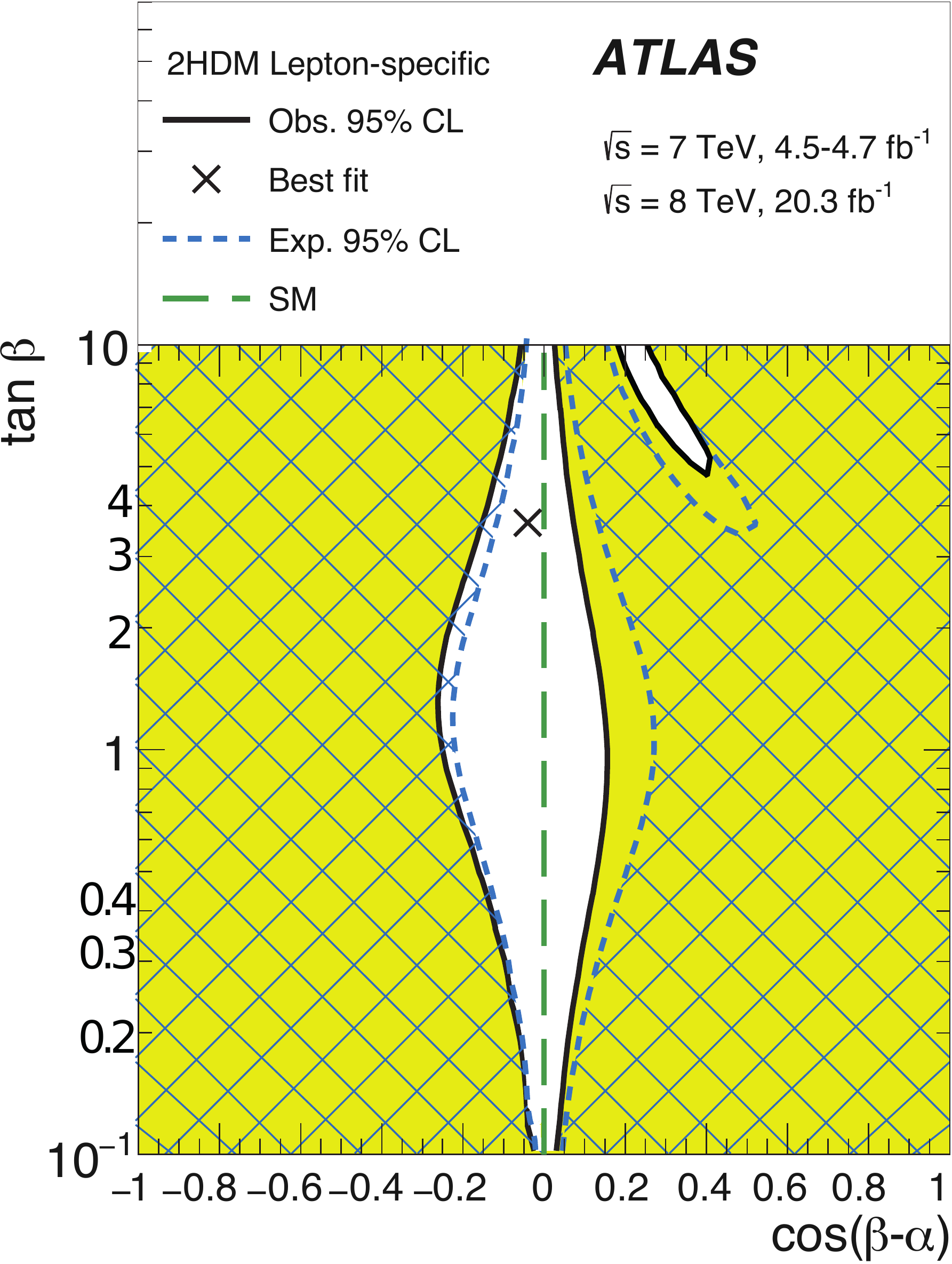}} \hspace{0.75cm}
\subfloat[Flipped]{\includegraphics[width=0.45\textwidth]{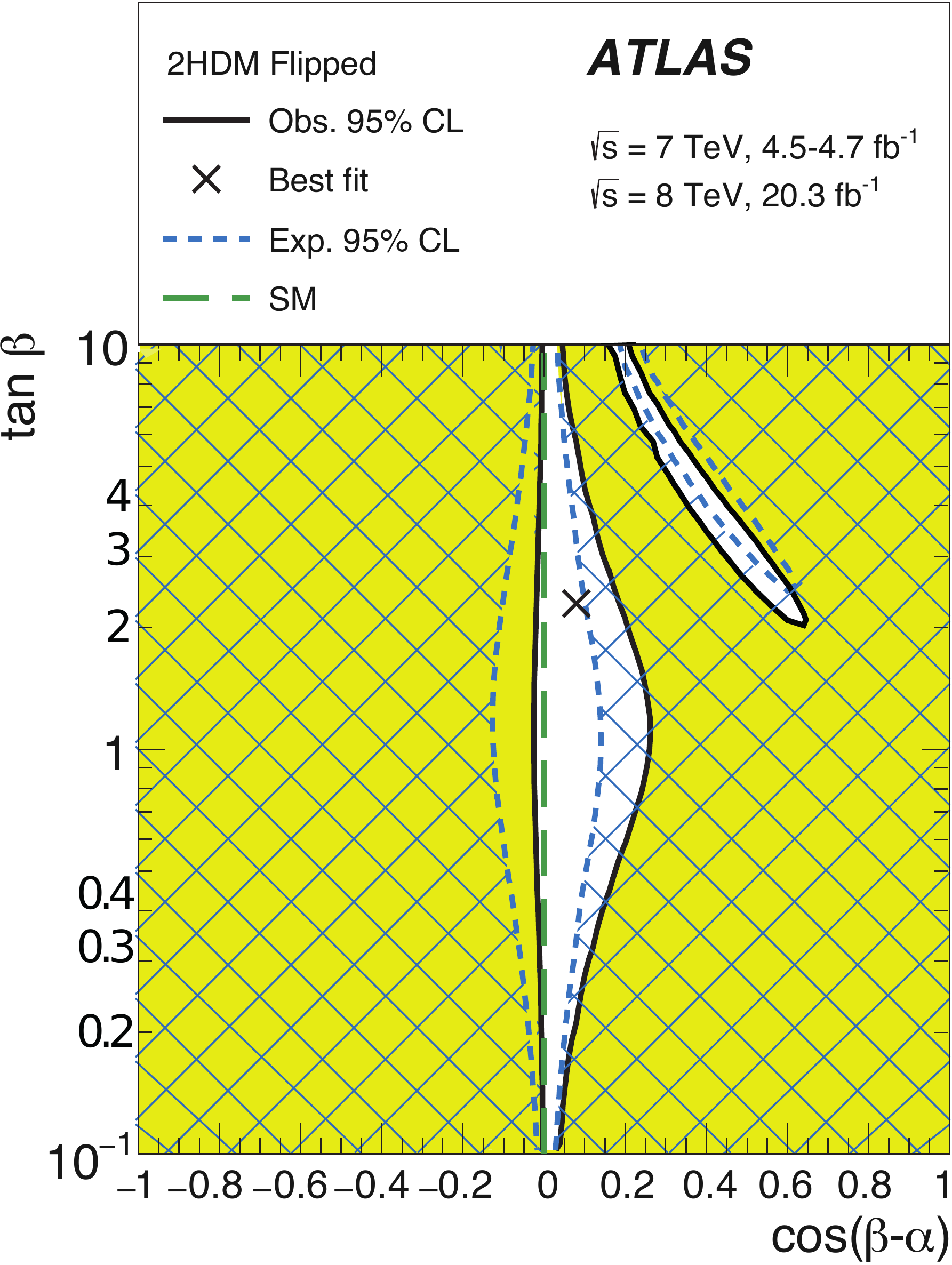}} \\
\caption{Regions of the $[\cos(\beta-\alpha),\ \tan\beta]$ plane of four types of 2HDMs excluded by fits to the measured rates of Higgs boson production and decays.  The likelihood contours where $-2\ln\Lambda=6.0$, corresponding approximately to the 95\% CL (2~std.~dev.), are indicated for both the data and the expectation for the SM Higgs sector.  The cross in each plot marks the observed best-fit value.  The light shaded and hashed regions indicate the observed and expected exclusions, respectively.  The $\alpha$ and $\beta$ parameters are taken to satisfy $0 \le \beta \le \pi/2$ and $0 \le \beta-\alpha \le \pi$ without loss of generality.}
\label{fig:2HDM_Overlay}
\end{center}
\end{figure*}

For this analysis, only the range 0.1$\le \tan\beta \le $10 was considered.  The regions of compatibility extend to larger and smaller $\tan\beta$ values, but with a correspondingly narrower range of $\cos(\beta-\alpha)$.  The confidence intervals drawn are derived from a \chisq distribution with two parameters of interest, corresponding to the quantities $\cos(\beta-\alpha)$ and $\tan\beta$.  However, at $\cos(\beta-\alpha)=0$ the likelihood is independent of the model parameter $\beta$, effectively reducing the number of parameters of interest locally to one.  Hence the test-statistic distribution for two parameters of interest that is used leads to some overcoverage near $\cos(\beta-\alpha)=0$.

}

\section{Simplified Minimal Supersymmetric Standard Model}{
\label{sec:SUSY}

Supersymmetry provides a means to solve the hierarchy problem by introducing superpartners of the corresponding SM particles.  Many supersymmetric models also provide a candidate for a dark-matter particle.
In the Minimal Supersymmetric Standard Model~\cite{Nilles:1983ge,Haber:1984rc,Gunion:1984yn,Barbieri:1987xf,Drees2004,Baer2006,Djouadi:2005gj}, the mass matrix of the neutral CP-even Higgs bosons $h$ and $H$ can be written as~\cite{Djouadi:2015jea}:

\begin{align}
\mathcal{M}_{\Phi}^{2} =
\begin{bmatrix}
m_{Z}^{2} \cos^{2}\beta + m_{A}^{2} \sin^{2}\beta & -(m_{Z}^{2}+m_{A}^2) \sin\beta \cos\beta \\
-(m_{Z}^{2}+m_{A}^{2})\sin\beta\cos\beta & m_{Z}^{2} \sin^{2}\beta + m_{A}^{2} \cos^{2}\beta
\end{bmatrix}
\nonumber & +
\begin{bmatrix}
\Delta\mathcal{M}_{11}^{2} & \Delta\mathcal{M}_{12}^{2} \\
\Delta\mathcal{M}_{12}^{2} & \Delta\mathcal{M}_{22}^{2}
\end{bmatrix} \quad ,
\end{align}
with radiative corrections being included through the 2$\times$2 matrix $\Delta \mathcal{M}_{ij}^{2}$.

A simplified approach to the study of the MSSM Higgs sector, known as the hMSSM~\cite{Maiani:2013hud, Djouadi:2013uqa, Djouadi:2015jea}, consists of neglecting the terms  $\Delta\mathcal{M}_{11}^{2}$ and $\Delta\mathcal{M}_{12}^{2}$.  The remaining term $\Delta\mathcal{M}_{22}^{2}$, which contains the dominant corrections from loops involving top quarks and stop squarks, is traded for the lightest mass eigenvalue $m_{h}$.
The scale factors for the Higgs boson couplings to vector bosons, up-type fermions, and down-type fermions ([$\kappa_{V}$, $\kappa_{u}$, $\kappa_{d}$]), can be expressed as functions of the free parameters [$m_A$, $\tan\beta$] (in addition to $m_{h}$) as~\cite{Maiani:2013hud, Djouadi:2013uqa, Djouadi:2015jea}:
\begin{equation}
\displaystyle
\begin{array}{ll}
\vspace{0.2cm}
\kappa_{V} = \scalebox{1.2}{$\frac{s_{d}(m_A, \tan\beta) + \tan\beta\,\,s_{u}(m_A, \tan\beta) }{\sqrt{1+\tan^2\beta}}$} \\
\vspace{0.2cm}
\kappa_{u} = s_{u}(m_A, \tan\beta) \scalebox{1.2}{$\frac{\sqrt{1+\tan^2\beta}}{\tan\beta}$} \\
\vspace{0.2cm}
\kappa_{d} = s_{d}(m_A, \tan\beta) \sqrt{1+\tan^2\beta} \quad , \\
\end{array}
\label{eqn:hMSSM}
\end{equation}
where the functions $s_{u}$ and $s_{d}$ are given by:
\begin{equation}
\displaystyle
\begin{array}{ll}
\vspace{0.2cm}
s_{u} = \scalebox{1.2}{$\frac{1}{\sqrt{1+\frac{\left(m_A^2 + m_Z^2\right)^2\,\tan^2\beta}{\left(m_Z^2 + m_A^2\,\tan^2\beta\,-\,m_h^2\left(1+\tan^2\beta\right)\right)^2} }}$} \\
\vspace{0.2cm}
s_{d} = \scalebox{1.2}{$\frac{\left(m_A^2\,+\,m_Z^2\right)\,\tan\beta}{m_Z^2\,+\,m_A^2\,\tan^2\beta\,-\,m_h^2\left(1+\tan^2\beta\right)}\,$}s_{u} \quad , \\
\end{array}
\label{eqn:su_sd}
\end{equation}
and $m_{Z}$ is the mass of the $Z$ boson.

To test the hMSSM model, the measured production and decay rates are expressed in terms of the corresponding coupling scale factors for vector bosons ($\kappa_{V}$), up-type fermions ($\kappa_{u}$), and down-type fermions ($\kappa_{d}$).  The observed Higgs boson is taken to be the light CP-even neutral Higgs boson $h$.  In the hMSSM, it is assumed to have the same production and decay modes as in the SM.  For comparison, Model~4 of Table~\ref{tab:Couplings} lists the measurements of ratios of the coupling scale factors [$\kappa_{V}$, $\kappa_{u}$, $\kappa_{d}$]~\cite{Aad:2015gba}.  The coupling scale factors are in turn cast in terms of $m_{A}$ and $\tan\beta$ using Eq.~(\ref{eqn:hMSSM}).  A correction is applied for $bbh$ associated production as a function of the $b$-quark Yukawa coupling as described in Section~\ref{sec:2HDM}.

Loop corrections from stops in ggF production, which can decrease the rate by 10--15\% for a light stop~\cite{Carena:2013qia}, and in diphoton decays are neglected.
Light tau sleptons (staus) with large mixing could enhance the diphoton rate by up to 30\% at $\tan\beta=50$~\cite{Carena:2013qia}, and charginos could modify the diphoton rate by up to 20\%~\cite{Nilles:1983ge, Haber:1984rc, Barbieri:1987xf, Bechtle:2013wla}; these effects are not included in the hMSSM model.

Additional corrections in the MSSM would break the universality of down-type fermion couplings, resulting for example in $\kappa_{b} \ne \kappa_{\tau}$.  These are generally sub-dominant effects~\cite{Maiani:2013hud, Djouadi:2013uqa, Djouadi:2015jea} and are not included.  The MSSM includes other possibilities such as Higgs boson decays to supersymmetric particles, decays of heavy Higgs bosons to lighter ones~\cite{Aad:2014yja}, and effects from light supersymmetric particles~\cite{Carena:2013qia}, which are not investigated here.  This model is therefore not fully general but serves as a useful benchmark, particularly if no direct observation of supersymmetry is made.

Contours of the two-dimensional likelihood in the $[m_{A}, \tan\beta]$ plane for the hMSSM model are shown in Figure~\ref{fig:SUSY}.  The data are consistent with the SM decoupling limit at large $m_{A}$.  The observed (expected) lower limit at the 95\% CL on the CP-odd Higgs boson mass is at least $m_{A}>370$~GeV (310~GeV) for $1\le\tan\beta\le50$, increasing to 440~GeV (330~GeV) at $\tan\beta=1$.  The observed limit is stronger than expected because the measured rates in the $h\to\gamma\gamma$~\cite{Aad:2014eha} (expected to be dominated by a $W$ boson loop in the SM) and $h\to ZZ^*\to 4\ell$~\cite{Aad:2014eva} channels are higher than predicted by the SM, but the hMSSM model has a physical boundary $\kappa_{V} \le 1$ so the vector-boson coupling cannot be larger than the SM value.  The physical boundary is accounted for by computing the profile likelihood ratio with respect to the maximum likelihood obtained within the physical region of the parameter space, $m_{A}>0$ and $\tan \beta>0$.  The region $1 \le \tan\beta \le 50$ is shown; at significantly smaller or larger values of $\tan\beta$, the hMSSM model is not a good approximation of the MSSM.  For $\tan\beta < 1$, the couplings to SM particles receive potentially large corrections related to the top sector that have not been included~\cite{Bagnaschi:2039911}.

The constraints in the $[m_{A},\ \tan\beta]$ plane of the hMSSM model from various direct searches for heavy Higgs bosons are also overlaid in Figure~\ref{fig:SUSY}.
The constraints from the following searches are shown.
\begin{itemize}
\item $H/A\to\tau\tau$ search via both ggF and $bbh$ associated production~\cite{Aad:2014vgg}.
\item Heavy CP-odd Higgs boson $A$ produced via ggF and decaying to $Zh$ with $Z\to ee$, $\mu\mu$, or $\nu\nu$ and $h\to bb$~\cite{Aad:2015wra}.
\item Heavy CP-even Higgs boson $H$ produced via ggF and decaying to $WW \to \ell\nu \ell\nu, ~\ell\nu qq$~\cite{Aad:2015agg} or $ZZ \to 4\ell,~\ell\ell qq,~\ell\ell bb,~\ell\ell\nu\nu$~\cite{Aad:2015kna} ($\ell = e$, $\mu$; $q = u,~d,~c,~s$).
\item Charged Higgs boson $H^{\pm}$ production in association with a top quark~\cite{Aad:2014kga}.
\end{itemize}

\begin{figure}[tbp]
\begin{center}
\includegraphics[width=0.95\textwidth]{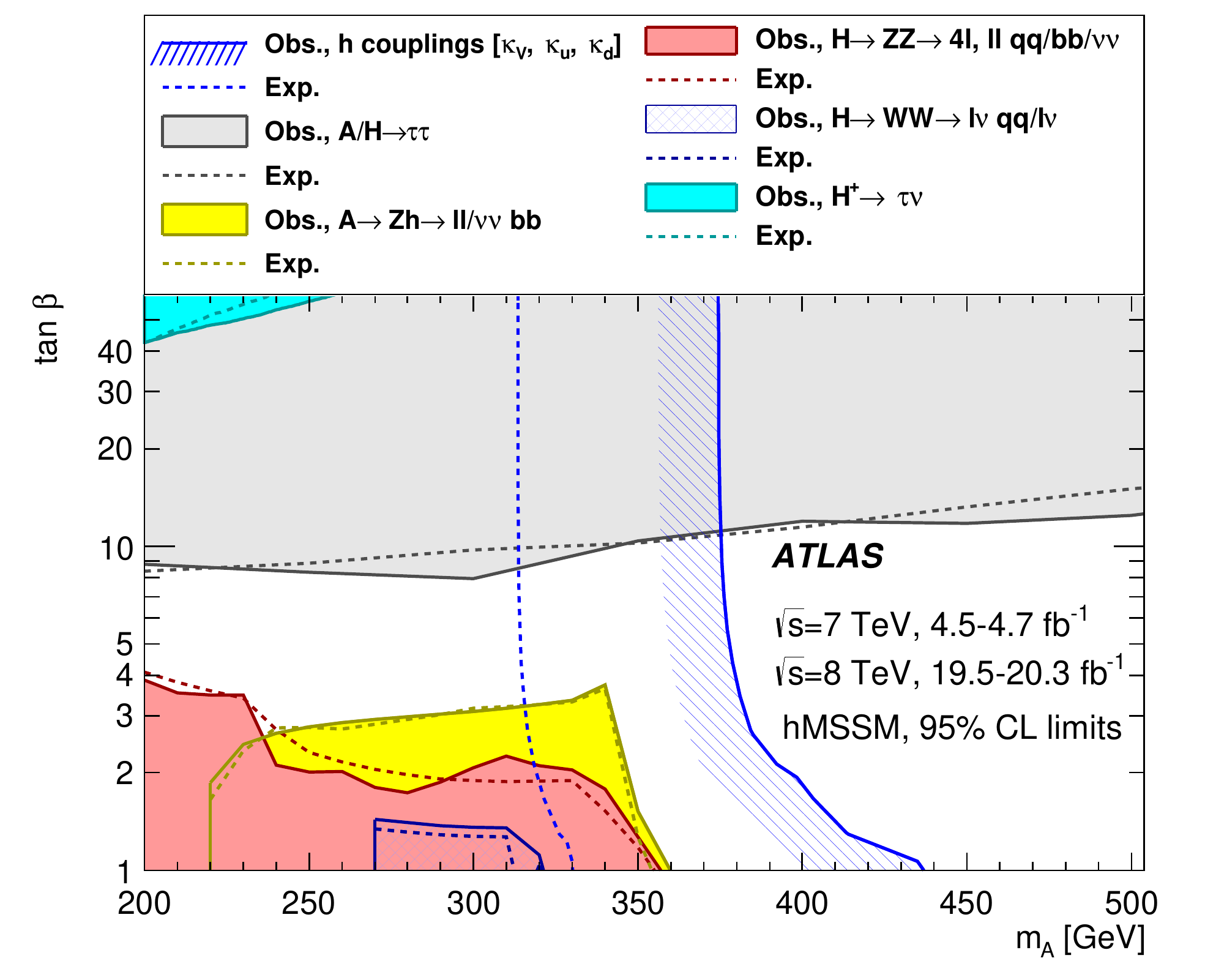}
\caption{Regions of the $[m_{A},\ \tan\beta]$ plane excluded in the hMSSM model via direct searches for heavy Higgs bosons and fits to the measured rates of observed Higgs boson production and decays.  The likelihood contours where $-2\ln\Lambda=6.0$, corresponding approximately to the 95\% CL (2~std.~dev.), are indicated for the data (solid lines) and the expectation for the SM Higgs sector (dashed lines).  The light shaded or hashed regions indicate the observed exclusions.  The SM decoupling limit is $m_{A} \to \infty$.}
\label{fig:SUSY}
\end{center}
\end{figure}

The cross sections for ggF and $bbh$ associated production in the five-flavour scheme hMSSM model have been calculated with \texttt{SUSHI~1.5.0}~\cite{Harlander:2012pb}.  The calculation for ggF includes the complete massive top and bottom loop corrections at next-to-leading-order (NLO) QCD~\cite{Harlander:2005rq}, the top quark loop corrections in the heavy-quark limit of QCD at next-to-next-to-leading-order (NNLO)~\cite{Harlander:2002wh, Anastasiou:2002yz, Ravindran:2003um}, and EW loop corrections due to light quarks up to NLO~\cite{Aglietti:2004nj, Actis:2008ug}.  The $bbh$ associated production in the five-flavour scheme includes corrections up to NNLO in QCD~\cite{Harlander:2003ai}.  The production of heavy Higgs bosons has also been calculated in the four-flavour scheme at NLO in QCD~\cite{Dittmaier:2003ej,Dawson:2003kb}, and the result has been combined with the five-flavour scheme using an empirical matching procedure~\cite{Harlander:2011aa}.  The branching ratios have been calculated using \texttt{HDECAY~6.4.2}~\cite{Djouadi:1997yw}.

}

\section{Probe of invisible Higgs boson decays} {
\label{sec:InvHiggsDecays}

\subsection{Direct searches for invisible decays} {
\label{sec:HInv}

Final states with large missing transverse momentum associated with leptons or jets offer the possibility of direct searches for $h\to$~invisible~\cite{Shrock:1982kd,Choudhury:1993hv,Eboli:2000ze,Davoudiasl:2004aj,Godbole:2003it,Ghosh:2012ep,Belanger:2013kya,Curtin:2013fra}.  In these searches, no excess of events was found and upper limits were set on the Higgs boson production cross section times the branching ratio for $h \to $~invisible decays. Assuming that the Higgs boson production cross sections and acceptances are unchanged relative to the SM expectations, upper bounds on the branching ratio of invisible Higgs boson decays, \BRinv, were obtained from the $\sigma \times \rm{BR}$ measurements. The ATLAS and CMS Collaborations set upper limits at the 95\% CL of 28\%~\cite{Aad:2015txa} and 65\%~\cite{Chatrchyan:2014tja}, respectively, on the branching ratio for invisible Higgs boson decays by searching for vector-boson fusion production of a Higgs boson that decays invisibly.  Using the \ZllhInv\ signature, weaker bounds were obtained by both ATLAS and CMS, giving upper limits of 75\%~\cite{Aad:2014iia} and 83\%~\cite{Chatrchyan:2014tja}, respectively. By combining searches in $Z(\ell\ell)h$ and $Z(bb)h$, CMS obtained an upper limit of 81\%~\cite{Chatrchyan:2014tja}.  A combination of the searches in VBF and $Zh$ production was carried out by CMS, giving a combined upper limit of 58\%~\cite{Chatrchyan:2014tja}.  Using the associated production with a vector boson, $Vh$, where $V=W$ or $Z$, $V\to jj$, and $h\to$~invisible, ATLAS set an upper bound of 78\%~\cite{Aad:2015uga}.  Other searches for invisible Higgs boson decays in events with large \ETmiss\ in association with one or more jets were also performed~\cite{Aad:2013oja,Khachatryan:2014rra,Aad:2015zva,monojetTheory}, but these searches are less sensitive to Higgs-mediated interactions. In the SM, the process \hZZnunununu\ is an invisible decay mode of the Higgs boson, but the branching ratio is 0.1\%~\cite{Heinemeyer:2013tqa}, which is below the sensitivities of the aforementioned direct searches.

A statistical combination of the following direct searches for invisible Higgs boson decays is performed:
\begin{enumerate}
\item[(1)] The Higgs boson is produced in the VBF process and decays invisibly~\cite{Aad:2015txa}. The signature of this process is two jets with a large separation in pseudorapidity, forming a large invariant dijet mass, together with large \ETmiss.
\item[(2)] The Higgs boson is produced in association with a $Z$ boson, where $Z\to \ell\ell$ and the Higgs boson decays to invisible particles~\cite{Aad:2014iia}. The signature in this search is two opposite-sign and same-flavour leptons (electrons or muons) with large missing transverse momentum. 
\item[(3)] The Higgs boson is produced in association with a vector boson $V$ ($W$ or $Z$), where $V\to jj$ and the Higgs boson decays to invisible particles~\cite{Aad:2015uga}. The signature in this search is two jets whose invariant mass $m_{jj}$ is consistent with the $V$ mass, together with large missing transverse momentum.
\end{enumerate}

To combine the measurements, the searches need to be performed in non-overlapping regions of phase space or the combination must account for the overlap in phase space.  The \ZllhInv\ search does not overlap with the other searches for $h\to$~invisible because a veto on events containing jets was required.  The overlap due to possible inefficiency in the veto requirements is negligible.  The \VBFhInv\ and the \VjjhInv\ searches also do not overlap in their phase spaces because the former requires a large dijet invariant mass (above $m_{jj} >500$~GeV) and latter imposes the requirement that the dijet invariant mass must be consistent with the associated vector boson mass within $50 < m_{jj} < 100$~GeV and imposes a veto on forward jets.  The same overlap removal requirements were applied in data to both the signal and control regions in the various searches, making the control regions used for background estimation non-overlapping.

The following nuisance parameters are treated as being fully correlated across the individual searches, with the rest being uncorrelated:
\begin{itemize}
\item Uncertainty in the luminosity measurements. This impacts the predicted rates of the signals and the backgrounds that are estimated using Monte Carlo simulation, namely ggF, VBF, and $Vh$ signals, and $t\bar{t}$, single top, and diboson backgrounds.
\item Uncertainties in the absolute scale of the jet energy calibration  and on the resolution of the jet energy calibration.
\item Uncertainties in the modelling of the parton shower.
\item Uncertainties in the renormalisation and factorisation scales, as well as the parton distribution functions. This affects the expected numbers of signal events in the ggF, VBF and $Vh$ production channels.
\end{itemize}
The uncertainty in the soft component of the missing transverse momentum has a significant impact in the \VjjhInv\ channel.  Its impact is much smaller in the other searches and not included as a nuisance parameter.  This uncertainty is therefore not correlated across all the searches. 

The limit on the branching ratio of $h\to$~invisible, defined in Eq.~(\ref{eq:invBR}), is computed assuming the SM production cross sections of the Higgs boson.  This is done using a maximum-likelihood fit to the event counts in the signal regions and the data control samples following the CL$_{S}$ modified frequentist formalism with a profile likelihood-ratio test statistic~\cite{Cowan:2010st}.

The statistical combination of direct searches for invisible Higgs boson decays results in an observed (expected) upper limit at the 95\% CL on \BRinv\ of 0.25~(0.27).  Figure~\ref{fig:combscan} shows the scan of the CL as a function of BR($h\to$~invisible) for the statistical combination of direct searches.  The limit obtained with the CL$_{S}$ method is consistent to two significant figures with the limit based on the log likelihood ratio.  Table~\ref{tab:comb} summarises the limits from direct searches for invisible Higgs boson decays and their statistical combination.  The combined limit is dominated by the \VBFhInv\ channel, which is by far the most sensitive.

\begin{figure*}[tbp]
\begin{center}
\includegraphics[width=0.70\textwidth]{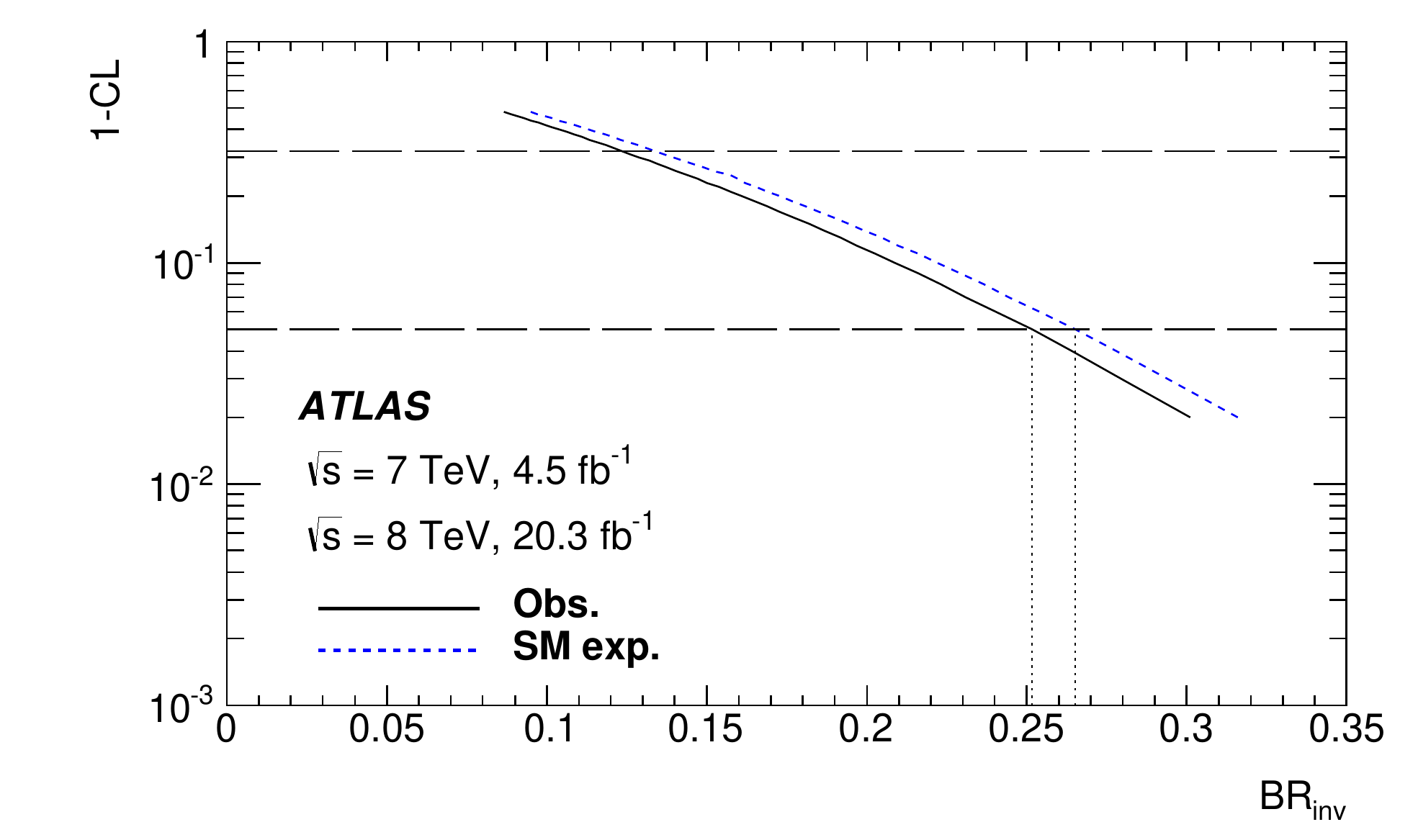}
\caption{The $(1-\rm{CL})$ versus BR($h\to$~invisible) scan for the combined search for invisible Higgs boson decays.  The horizontal dashed lines refer to the 68\% and 95\% confidence levels. The vertical dashed lines indicate the observed and expected upper bounds at the 95\% CL on BR($h\to$~invisible) for the combined search.}
\label{fig:combscan}
\end{center}
\end{figure*}

\begin{table}[tbp]
\begin{center}
\begin{tabular}{ c | c c c c c c }
\hline
\hline
Channels & \multicolumn{6}{c}{Upper limit on BR($h\rightarrow$~inv.) at the 95\% CL} \\
& Obs.   & $-2$~std.~dev.   &  $-1$~std.~dev. & Exp.      & +1~std.~dev.   & +2~std.~dev.  \\
\hline
\hline
VBF $h$                & 0.28      & 0.17         & 0.23        & 0.31          & 0.44         & 0.60             \\
$Z(\to \ell\ell)h$     & 0.75       & 0.33         & 0.45        & 0.62          & 0.86         & 1.19              \\
$V(\to jj)h$           & 0.78       & 0.46         & 0.62        & 0.86          & 1.19         & 1.60             \\
\hline
Combined Results       & 0.25       & 0.14         & 0.19        & 0.27          & 0.37         & 0.50              \\
\hline
\hline
\end{tabular}
\caption{Summary of upper bounds on BR($h\rightarrow$~invisible) at the 95\% CL from the individual searches and their combination.  The Higgs boson production rates via VBF and $Vh$ associated production are assumed to be equal to their SM values.  The numerical bounds larger than 1 can be interpreted as
an upper bound on $\sigma/\sigma_{{\rm SM}}$, where $\sigma_{{\rm SM}}$ is the Higgs boson production cross section in the SM. 
\label{tab:comb}}
\end{center}
\end{table}

}
\subsection{Combination of visible and invisible decay channels} {
\label{sec:VisibleInvisible}

The measurements of Higgs boson visible decay rates are complementary to the direct searches for invisible decays since they are indirectly sensitive to undetectable decays as well.  The visible decay rates are used to extract the sum of the branching fractions to invisible and undetectable final states.  A conservative limit on the invisible branching ratio is then inferred by assuming the undetectable branching ratio to be negligibly different from the SM expectation of approximately zero.  For example, a significant excess in a search for Higgs boson decays to a new particle would further tighten the indirect limit on the invisible branching ratio.  It would not affect the limits from direct searches for invisible Higgs boson decays.

The overall upper limit on the branching ratio of the Higgs boson to invisible final states, \BRinv, is derived using a statistical combination of measurements from both the visible and invisible Higgs boson decays.  The visible decay channels are \hgg, \hZZllll, \hWWlnln, \hZg, \htt, \hmm, and \hbb, with a variety of production mode selections used.  The invisible decay channels are described in Section~\ref{sec:HInv} and involve the Higgs boson being produced via VBF or $Z(\ell\ell)h$, and then decaying invisibly.  The $V(jj)h$ production mode is not included due to overlap of the event selection with the 0-lepton category of the $Vh(bb)$ measurement.

The extraction of \BRinv\ is performed using a coupling parameterisation that includes separate scale factors for the couplings of the Higgs boson to the $W$ boson, $Z$ boson, top quark, bottom quark, tau~lepton, and muon, as well as scale factors for effective loop-induced couplings to gluons, photons, and $Z\gamma$ to absorb the possible contributions of new particles through loops.  The Higgs boson production modes are taken to be the same as those in the SM.

As for the visible decay rates alone, the invisible branching ratio is conservatively estimated by taking the undetectable branching ratio to be zero.  Thus the scale factor $\kappa_{h}^2$ is equal to the ratio of the total width of the Higgs boson to the SM expectation, $\Gamma_{h} / \Gamma_{h, {\rm SM}}$, and can be expressed in terms of \BRinv\ as:
\begin{equation}
\begin{array}{ll}
\vspace{0.2cm}
\kappa_{h}^2 = \Gamma_{h} / \Gamma_{h, {\rm SM}} = \sum\limits_{j} \kappa_{j}^2 {\rm BR}_{j} / (1 - \BRinv) \quad .\\
\end{array}
\end{equation}

The production and decay rates of all channels are fit with functions of [$\kappa_{W}$, $\kappa_{Z}$, $\kappa_{t}$, $\kappa_{b}$, $\kappa_{\tau}$, $\kappa_{\mu}$, $\kappa_{g}$, $\kappa_{\gamma}$, $\kappa_{Z\gamma}$, \BRinv].  Each parameter of interest is determined by treating the others as nuisance parameters.

With the visible decay channels alone, the assumption that the vector-boson coupling is less than or equal to the SM expectation ($\kappa_{V}\le1$) produced an observed (expected) upper limit at the 95\% CL of $\BRinv <$ 0.49 (0.48)~\cite{Aad:2015gba}.  With only the invisible decay channels (including $V(jj)h$), the assumption that the vector-boson and gluon couplings are identical to their SM expectations ($\kappa_{i}=1$) yields an upper limit at the 95\% CL of 0.25 (0.27), as described in Section~\ref{sec:HInv}.
 
When the visible and invisible decay channels (but excluding $V(jj)h$) are combined, no assumption about $\kappa_{V}$ or other couplings is made beyond that on the undetectable branching ratio.  The resulting observed likelihood scan as a function of \BRinv\ is shown in Figure~\ref{fig:VisInv}.  The fitted values of the Higgs boson couplings are similar to those in the SM, so the measured overall signal strength($\mu_{h}>$1) is accommodated in the fit by a best-fit value of \BRinv\ that would be negative.  This decreases the Higgs boson total width, which is inversely proportional to the signal strength.  Scans of the observed likelihoods of the invisible and visible decay channels separately are also shown.

\begin{figure*}[tbp]
\begin{center}
\includegraphics[width=0.6\textwidth]{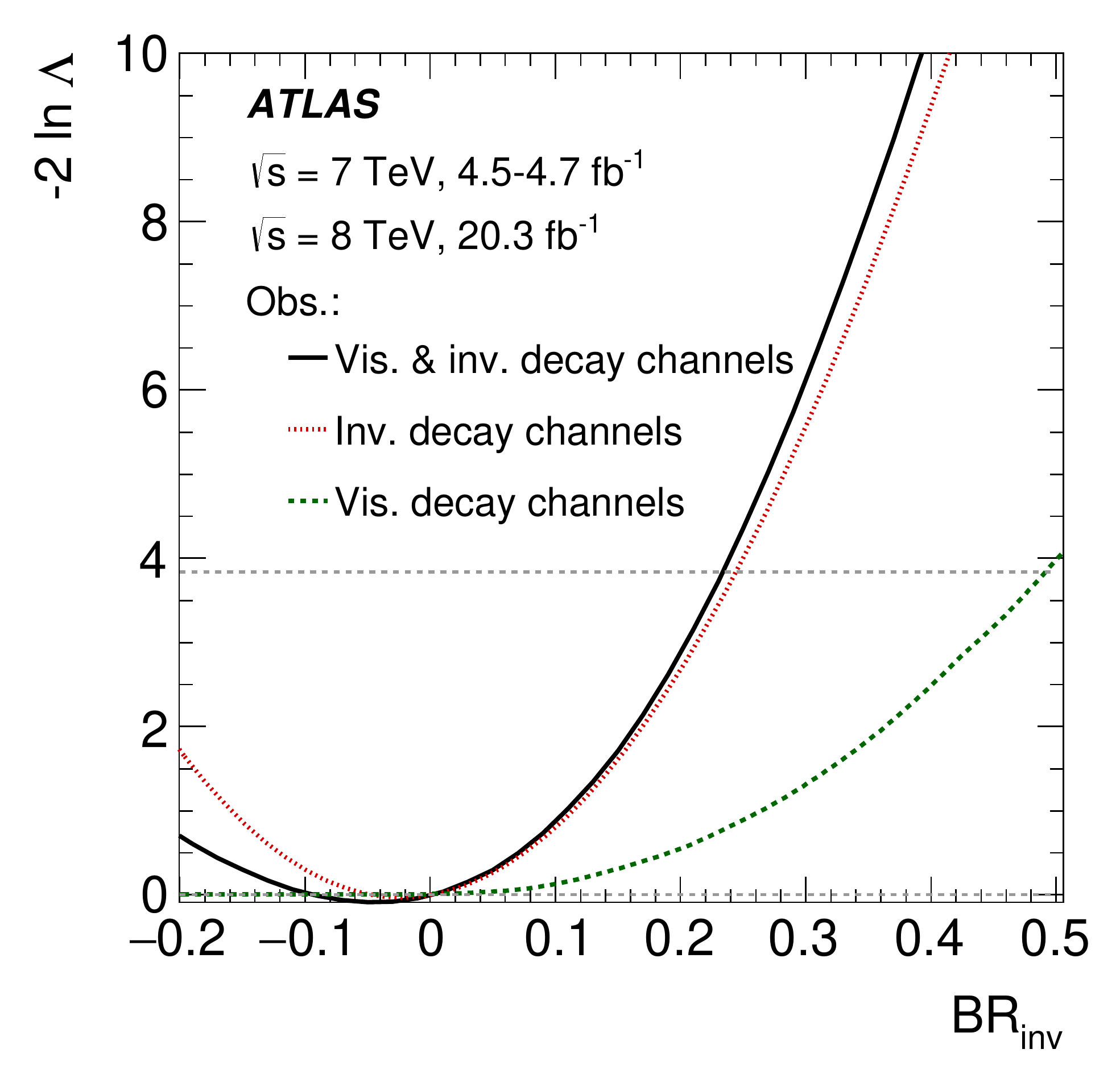}
\caption{Observed likelihood scans of the Higgs boson invisible decay branching ratio using direct searches for invisible Higgs boson decays, rate measurements in visible Higgs boson decays, and the overall combination of invisible and visible decay channels.  The line at $-2\ln\Lambda=0$ corresponds to the most likely value of \BRinv\ within the physical region $\BRinv \ge 0$.  The line at $-2\ln\Lambda=3.84$ corresponds to the one-sided upper limit at approximately the 95\% CL (2~std.~dev.), given $\BRinv \ge 0$.}
\label{fig:VisInv}
\end{center}
\end{figure*}

The observed (expected) upper limit at the 95\% CL is $\BRinv <$ 0.23 (0.24) using the combination of all channels and accounting for the physical boundary.\footnote{The observed upper limit at the 90\% CL is $\BRinv <$ 0.22, which is used in Section~\ref{sec:DarkMatter}.}  This baseline configuration, with a maximally general set of independent coupling parameters and no explicit assumption about the value of $\kappa_{W,Z}$, provides the main result for the combination of invisible and visible decay channels.  The results are dominated by the direct searches for invisible decays.  The expected limit demonstrates a fractional improvement of 11\% in sensitivity compared to the invisible decay channels alone.  In addition, it is more model-independent than the combination of invisible (visible) decay channels alone, because it does not assume the vector-boson couplings to be equal to (less than or equal to) their SM values.  The absolute couplings to the $Z$ boson, $W$ boson, top quark, bottom quark, tau lepton, muon, gluon, photon, and $Z\gamma$ are extracted in the baseline scenario.  The measurements or limits are given, along with the upper limit on \BRinv, in Model~6 of Table~\ref{tab:Couplings}.

\begin{table}[tbp]
\begin{center}
\begin{tabular}{ c | c | c | c c }
\hline
\hline
Decay channels & Coupling parameterisation & $\kappa_{i}$ assumption & \multicolumn{2}{c}{Upper limit on \BRinv} \\
& & & Obs. & Exp. \\
\hline
\hline
Invisible decays & [$\kappa_{W}$, $\kappa_{Z}$, $\kappa_{t}$, $\kappa_{b}$, $\kappa_{\tau}$, $\kappa_{\mu}$, $\kappa_{g}$ $\kappa_{\gamma}$, $\kappa_{Z\gamma}$, \BRinv] & $\kappa_{W,Z,g} = 1$ & 0.25 & 0.27 \\
Visible decays & [$\kappa_{W}$, $\kappa_{Z}$, $\kappa_{t}$, $\kappa_{b}$, $\kappa_{\tau}$, $\kappa_{\mu}$, $\kappa_{g}$ $\kappa_{\gamma}$, $\kappa_{Z\gamma}$, \BRinv] & $\kappa_{W,Z}\le 1$ & 0.49 & 0.48 \\
\hline
\bf{Inv. \& vis. decays} & \bf{[$\kappa_{W}$, $\kappa_{Z}$, $\kappa_{t}$, $\kappa_{b}$, $\kappa_{\tau}$, $\kappa_{\mu}$, $\kappa_{g}$ $\kappa_{\gamma}$, $\kappa_{Z\gamma}$, \BRinv]} & \bf{None} & \bf{0.23} & \bf{0.24} \\
Inv. \& vis. decays & [$\kappa_{W}$, $\kappa_{Z}$, $\kappa_{t}$, $\kappa_{b}$, $\kappa_{\tau}$, $\kappa_{\mu}$, $\kappa_{g}$ $\kappa_{\gamma}$, $\kappa_{Z\gamma}$, \BRinv] & $\kappa_{W,Z}\le 1$ & 0.23 & 0.23 \\
\hline
\hline
\end{tabular}
\caption{Summary of upper limits on BR($h\rightarrow$~invisible) at the 95\% CL from the combination of direct searches for invisible Higgs boson decays, the combination of measurements of visible Higgs boson decays, and the overall combination using both the invisible and visible Higgs boson decays.  The results are derived using different assumptions about $\kappa_{W,Z}$.  The results with the baseline configuration for the combination of invisible and visible decay channels are indicated in bold.
\label{tab:VisInv}}
\end{center}
\end{table}

As an alternate scenario, if the assumption $\kappa_{W,Z}\le1$ is added for the combination of channels, the observed (expected) limit is 0.23 (0.23).  This allows a useful comparison with the results of the invisible or visible decay channels alone, which apply a similar assumption.  The results from the invisible and visible decay channels separately, as well as their statistical combination, are summarised in Table~\ref{tab:VisInv} for the coupling parameterisation and assumptions about $\kappa_{W,Z}$ used.  The baseline results for the combination of invisible and visible decays are highlighted.

A less general coupling parameterisation [$\kappa_{V}$, $\kappa_{F}$, $\kappa_{g}$, $\kappa_{\gamma}$, $\kappa_{Z\gamma}$, \BRinv] was also considered.  The reduction in the number of degrees of freedom would make the limit significantly more model-dependent and only improve the sensitivity marginally, so this parametrisation was not used.

}
\subsection{Higgs portal to dark matter}{
\label{sec:DarkMatter}

Many ``Higgs portal'' models~\cite{Shrock:1982kd, Patt:2006fw, Kanemura:2010sh, Fox:2011pm, Djouadi:2011aa, LopezHonorez:2012kv, Aad:2014iia, Aad:2013oja} introduce an additional weakly interacting massive particle (WIMP) as a dark-matter candidate.  It is taken to interact very weakly with the SM particles, except for the Higgs boson.  In this study, the coupling of the Higgs boson to the WIMP is taken to be a free parameter.  The Higgs portal model, where the Higgs boson is taken to be the only mediator, is a special case of the spin-independent limits obtained by direct detection experiments.

To compare with these direct searches, the observed upper limit combining all the visible and invisible Higgs boson decay channels using the most general baseline parameterisation is calculated at the 90\% CL:  this gives $\BRinv<0.22$.  This limit is translated into constraints on the coupling of the WIMP to the Higgs boson as a function of its mass~\cite{Djouadi:2011aa}.  This is done for WIMP masses less than half the Higgs boson mass, under the assumption that the resulting Higgs boson decays to WIMP pairs account entirely for \BRinv.  Any additional contributions to \BRinv\ from other new phenomena would produce a more stringent limit.  The partial width for Higgs boson decays to a pair of dark-matter particles depends on the spin of the dark-matter particle.  It is given for scalar, Majorana fermion, or vector dark-matter candidates (where the Majorana fermion is motivated by neutralinos in supersymmetry) as:
\begin{equation}
\begin{array}{ll}
{\rm scalar}\ S:  \Gamma^{\rm inv}(h\to SS)  = & \lambda^2_{hSS} \cfrac{v^2\beta_S}{128\pi m_h} \\
{\rm fermion}\ f: \Gamma^{\rm inv}(h\to ff) = & \cfrac{\lambda^2_{hff}}{\Lambda^2} \cfrac{v^2\beta^3_fm_h}{64\pi} \\
{\rm vector}\ V:  \Gamma^{\rm inv}(h\to VV)  = & \lambda^2_{hVV} \cfrac{v^2\beta_V m_h^3}{512\pi m_V^4} \, \times \, \left(1-4\cfrac{m_V^2}{m_h^2}+12\cfrac{m_V^4}{m_h^4}\right) \quad . 
\end{array}
\label{eq:GammaInv}
\end{equation}
\noindent
Here $\lambda_{hSS}$, $\lambda_{hff} / \Lambda$, and $\lambda_{hVV}$ are the couplings of the Higgs boson to dark-matter particles of corresponding spin, $v$ denotes the vacuum expectation value of the Higgs boson, and $\beta_\chi = \sqrt{1-4m_{\chi}^2 / m_h^2}$ is a kinematic factor associated with the two-body $h\to \chi\chi$ decay ($\chi=S,\ V$, or $f$).  For the cases of a fermion or vector boson WIMP in this effective field theory approach, the new physics scale $\Lambda$ is assumed to be at the TeV scale or higher, well above the probed scale at the SM Higgs boson mass.  These equations are used to deduce the coupling of the Higgs boson to the WIMP for each of the three possible WIMP spins.  The coupling is then re-parameterised in terms of the cross section for scattering between the WIMP and nucleons via Higgs boson exchange, $\sigma_{\chi-N}$~\cite{Djouadi:2011aa}:
\begin{equation}
\begin{array}{ll}
  \vspace{0.2cm}
  {\rm scalar}\ S: & \sigma_{S-N} = \lambda^2_{hSS} \cfrac{m_N^4 f_N^2}{16\pi m_h^4 (m_S+m_N)^2} \\
  \vspace{0.2cm}
  {\rm fermion}\ f: & \sigma_{f-N} = \cfrac{\lambda^2_{hff}}{\Lambda^2} \cfrac{m_N^4 f_N^2 m_f^2}{4\pi m_h^4 (m_f+m_N)^2} \\
  {\rm vector}\ V: & \sigma_{V-N} = \lambda^2_{hVV} \cfrac{m_N^4 f_N^2}{16\pi m_h^4 (m_V+m_N)^2} \quad ,
\end{array}
\label{eq:SigmaChiN}
\end{equation}
where $m_N \sim 0.94$~GeV is the nucleon mass, and $f_N = 0.33^{+0.30}_{-0.07}$ is the form factor associated with the Higgs boson--nucleon coupling, computed using lattice QCD~\cite{Djouadi:2011aa}.  Upper limits at the 90\% CL on the WIMP--nucleon scattering cross section $\sigma_{\chi-N}$ are derived as a function of the WIMP mass $m_\chi$, as shown in Figure~\ref{fig:HiggsPortal}.  The hashed bands indicate the uncertainty resulting from the systematic variation of the form factor $f_N$.  They are compared with limits from direct searches for dark matter~\cite{Angle:2011th, Aprile:2012nq, Bernabei:2008yi, Angloher:2011uu, Agnese:2013rvf, Aalseth:2011wp, Akerib:2013tjd, Agnese:2014aze, Angloher:2014myn} at the confidence levels indicated.  The ATLAS limits on the WIMP--nucleon scattering cross section are proportional to those on the invisible decay branching ratio, as evident from Eqns.~(\ref{eq:GammaInv})--(\ref{eq:SigmaChiN}).  They are weaker (stronger) at low mass for scalar (Majorana and vector) WIMPs, and degrade as $m_\chi$ approaches $m_h/2$ as expected from kinematics.  The limits are shown for $m_\chi \ge 1$~GeV, but extend to WIMP masses smaller than this value.

\begin{figure*}[tbp]
\begin{center}
\includegraphics[width=\columnwidth]{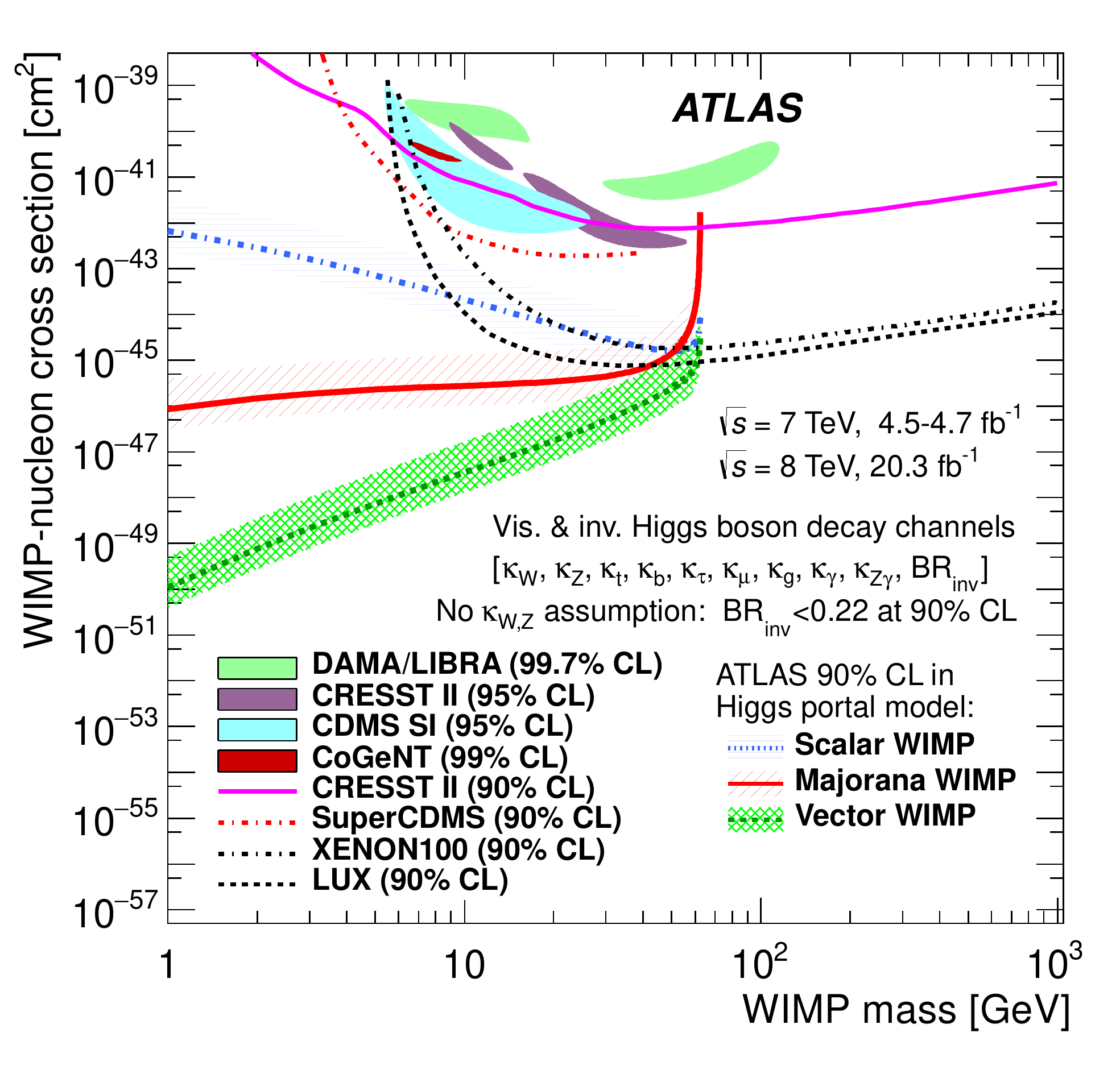}
\caption{ATLAS upper limit at the 90\% CL on the WIMP--nucleon scattering cross section in a Higgs portal model as a function of the mass of the dark-matter particle, shown separately for a scalar, Majorana fermion, or vector-boson WIMP.  It is determined using the limit at the 90\% CL of $\BRinv<0.22$ derived using both the visible and invisible Higgs boson decay channels.  The hashed bands indicate the uncertainty resulting from varying the form factor $f_N$ by its uncertainty.  Excluded and allowed regions from direct detection experiments at the confidence levels indicated are also shown~\cite{Angle:2011th, Aprile:2012nq, Bernabei:2008yi, Angloher:2011uu, Agnese:2013rvf, Aalseth:2011wp, Akerib:2013tjd, Agnese:2014aze, Angloher:2014myn}.  These are spin-independent results obtained directly from searches for nuclei recoils from elastic scattering of WIMPs, rather than being inferred indirectly through Higgs boson exchange in the Higgs portal model.}
\label{fig:HiggsPortal}
\end{center}
\end{figure*}
}

}

\section{Conclusions} {
\label{sec:Conclusions}

Higgs boson coupling measurements from the combination of multiple production and decay channels (\hgg, \hZZllll, \hWWlnln, \hZg, \hbb, \htt, \hmm) have been used to indirectly search for new physics.  The results are based on up to 4.7~fb$^{-1}$ of $pp$ collision data at $\sqrt{s}=7$~TeV and 20.3~fb$^{-1}$ at $\sqrt{s}=8$~TeV recorded by the ATLAS experiment at the LHC.  No significant deviation from the SM expectation is found in the observables studied, which are used to constrain various models of new phenomena.

The mass dependence of the couplings is consistent with the predictions for a SM Higgs boson.  Constraints are set on Minimal Composite Higgs Models, models with an additional electroweak singlet, and two-Higgs-doublet models.  A lower limit at the 95\%~CL is set on the mass of a pseudoscalar Higgs boson in the hMSSM of 370~GeV.  Results from direct searches for heavy Higgs bosons are also interpreted in the hMSSM.  In addition, direct searches for invisible Higgs boson decays in the VBF, $Z(\ell\ell)h$, and $V(jj)h$ production modes are combined to set an upper bound at the 95\% CL on the Higgs boson invisible decay branching ratio of 0.25.  Including the coupling measurements in visible decays further improves the upper limit to 0.23.  The limit on the invisible decay branching ratio is used to constrain the rate of dark matter--nucleon scattering in a model with a Higgs portal to dark matter.
}

% Acknowledgements for papers with collision data
% Version 15-Sep-2015

% Standard acknowledgements start here
%----------------------------------------------
We thank CERN for the very successful operation of the LHC, as well as the
support staff from our institutions without whom ATLAS could not be
operated efficiently.

We acknowledge the support of ANPCyT, Argentina; YerPhI, Armenia; ARC, Australia; BMWFW and FWF, Austria; ANAS, Azerbaijan; SSTC, Belarus; CNPq and FAPESP, Brazil; NSERC, NRC and CFI, Canada; CERN; CONICYT, Chile; CAS, MOST and NSFC, China; COLCIENCIAS, Colombia; MSMT CR, MPO CR and VSC CR, Czech Republic; DNRF, DNSRC and Lundbeck Foundation, Denmark; IN2P3-CNRS, CEA-DSM/IRFU, France; GNSF, Georgia; BMBF, HGF, and MPG, Germany; GSRT, Greece; RGC, Hong Kong SAR, China; ISF, I-CORE and Benoziyo Center, Israel; INFN, Italy; MEXT and JSPS, Japan; CNRST, Morocco; FOM and NWO, Netherlands; RCN, Norway; MNiSW and NCN, Poland; FCT, Portugal; MNE/IFA, Romania; MES of Russia and NRC KI, Russian Federation; JINR; MESTD, Serbia; MSSR, Slovakia; ARRS and MIZ\v{S}, Slovenia; DST/NRF, South Africa; MINECO, Spain; SRC and Wallenberg Foundation, Sweden; SERI, SNSF and Cantons of Bern and Geneva, Switzerland; MOST, Taiwan; TAEK, Turkey; STFC, United Kingdom; DOE and NSF, United States of America. In addition, individual groups and members have received support from BCKDF, the Canada Council, CANARIE, CRC, Compute Canada, FQRNT, and the Ontario Innovation Trust, Canada; EPLANET, ERC, FP7, Horizon 2020 and Marie Skłodowska-Curie Actions, European Union; Investissements d'Avenir Labex and Idex, ANR, Region Auvergne and Fondation Partager le Savoir, France; DFG and AvH Foundation, Germany; Herakleitos, Thales and Aristeia programmes co-financed by EU-ESF and the Greek NSRF; BSF, GIF and Minerva, Israel; BRF, Norway; the Royal Society and Leverhulme Trust, United Kingdom.

The crucial computing support from all WLCG partners is acknowledged
gratefully, in particular from CERN and the ATLAS Tier-1 facilities at
TRIUMF (Canada), NDGF (Denmark, Norway, Sweden), CC-IN2P3 (France),
KIT/GridKA (Germany), INFN-CNAF (Italy), NL-T1 (Netherlands), PIC (Spain),
ASGC (Taiwan), RAL (UK) and BNL (USA) and in the Tier-2 facilities
worldwide.
%----------------------------------------------

\clearpage
\printbibliography
\newpage

% ATLAS Collaboration author list
% Data extracted on 11-Aug-2015 for paper reference HIGG-2015-03

%% \documentclass[11pt]{article}
%% \usepackage{a4wide}
%% \begin{document}

\begin{flushleft}
{\Large The ATLAS Collaboration}

\bigskip

G.~Aad$^{\rm 85}$,
B.~Abbott$^{\rm 113}$,
J.~Abdallah$^{\rm 151}$,
O.~Abdinov$^{\rm 11}$,
R.~Aben$^{\rm 107}$,
M.~Abolins$^{\rm 90}$,
O.S.~AbouZeid$^{\rm 158}$,
H.~Abramowicz$^{\rm 153}$,
H.~Abreu$^{\rm 152}$,
R.~Abreu$^{\rm 116}$
% ...
\bigskip
\\
$^{1}$ Department of Physics, University of Adelaide, Adelaide, Australia\\
% ...
$^{11}$ Institute of Physics, Azerbaijan Academy of Sciences, Baku, Azerbaijan\\
% ...
$^{85}$ CPPM, Aix-Marseille Universit{\'e} and CNRS/IN2P3, Marseille, France\\
% ...
$^{90}$ Department of Physics and Astronomy, Michigan State University, East Lansing MI, United States of America\\
% ...
$^{107}$ Nikhef National Institute for Subatomic Physics and University of Amsterdam, Amsterdam, Netherlands\\
% ...
$^{113}$ Homer L. Dodge Department of Physics and Astronomy, University of Oklahoma, Norman OK, United States of America\\
% ...
$^{116}$ Center for High Energy Physics, University of Oregon, Eugene OR, United States of America\\
% ...
$^{151}$ Institute of Physics, Academia Sinica, Taipei, Taiwan\\
$^{152}$ Department of Physics, Technion: Israel Institute of Technology, Haifa, Israel\\
$^{153}$ Raymond and Beverly Sackler School of Physics and Astronomy, Tel Aviv University, Tel Aviv, Israel\\
% ...
$^{158}$ Department of Physics, University of Toronto, Toronto ON, Canada\\
% ...
$^{s}$ Also at Institute of Theoretical Physics, Ilia State University, Tbilisi, Georgia\\
% ...
\end{flushleft}

%\end{document}
% Created with ./xml2latex.py

\end{document}